\documentclass[twocolumns]{aa}
\usepackage[varg]{txfonts}
\usepackage[english]{babel}
\usepackage{natbib}
\bibpunct{(}{)}{;}{a}{}{,}
\usepackage{hyperref}
\usepackage{color}

\usepackage{multirow}
\newcommand{\sgra}{\ensuremath{\text{Sgr\,A}^{\star} }}
\newcommand{\Msun}{\ensuremath{\text{M}_{\odot}}}
\newcommand{\Ka}{K$\alpha$~}
\usepackage{graphicx}
\usepackage{cprotect}

\begin{document}

   \title{Iron \Ka echoes from the circumnuclear disk orbiting \sgra}

   \subtitle{}

\author{Giovanni Stel\inst{1,2}\thanks{email: giovanni.stel@inaf.it}
        \and Gabriele Ponti\inst{2,3} 
        \and Francesco Haardt\inst{1,2,4}
        }

\institute{
        DiSAT, Universit\`a degli Studi dell'Insubria, via Valleggio 11, I-22100 Como, Italy
\and
        INAF -- Osservatorio Astronomico di Brera, Via E. Bianchi 46, 23807 Merate, Italy
\and 
        Max Planck Institute fur Extraterrestriche Physik, 85748, Garching, Germany
\and
        INFN, Sezione Milano-Bicocca, P.za della Scienza 3, I-20126 Milano, Italy}

   \date{}

 
  \abstract
   {Molecular clouds in the Galactic center (GC) reprocess radiation from past outbursts of nearby high-energy sources, generating a bright Fe \Ka fluorescence at 6.4 keV. The closest clouds to the GC are only $\simeq 1.5$ pc from \sgra, forming a torus-like structure known as the circumnuclear disk (CND). The study of fluorescence emission can lead to a characterization of the
illuminating source(s), the reflecting clouds, and the global geometry of such a system lying in the GC.}
   {The primary purpose of our study is to analyze possible fluorescence signals arising in the CND. This signal would allow us to constrain the CND's physical properties and the source-reflector system's geometry. }
   {By exploiting the last $\simeq 20$ years of XMM-Newton observations of the GC, we studied the variability of the Fe \Ka line in the region around \sgra. We identified regions with a flux excess and computed the spectrum therein. We then derived the hydrogen column density of the CND after relating the intensity of the 6.4 keV line to the total energy emitted by known transient sources in the region.
}
   {Starting from data collected in 2019, we find significant line excesses in a region compatible with the eastern portion of the CND. The echo radiation can be linked to the 2013 outburst of the magnetar SGR J1745-2900. We derive a mean effective hydrogen column density of the CND in the eastern region of $\simeq 10^{23}$ cm$^{-2}$. 
} 
   {The scenario depicted is physically plausible, given the luminosity, the position of the illuminating source, and the expected density of the CND. Further observations could link the variability of the echo signal to the light curve of the illuminating source. In this way, it would be possible to characterize the cloud response to the radiation front, achieving a more accurate estimate of the cloud parameters.}

   \maketitle

\section{Introduction}

The Galactic center (GC) is the most accessible environment for studying astrophysical processes around supermassive black holes \citep[SMBHs,][]{Genzel+2010}. Giant molecular clouds (GMCs) lie in the inner $\simeq$ 200 pc of the Galaxy in the so-called central molecular zone \citep[CMZ,][]{Morris+1996}. These clouds trace the past activity of bright sources inside the CMZ \citep{Sunyaev+1993,Markevitch+1993,Koyama+1996, Ponti+2013}. Indeed, GMCs show a strong Fe \Ka fluorescent line at 6.4 keV due to reprocessed radiation that has invested the cloud in the past \citep{Sunyaev+1998}. The Fe \Ka line emission from GMCs has been largely investigated in the literature, providing constraints on the past luminosity of \sgra, and suggesting that \sgra\  experienced past periods of much higher activity than today \citep[see, e.g.,][]{Koyama+1996,Muno+2007, Ponti+2010, Inui+2009, Nobukawa+2011,Clavel+2013,Zhang+2015,Chrazov+2017,Terrier+2018,Chuard+2018,Marin+2023}. \citet{Muno+2005} reported the observation of echo radiation from the outburst of the low-mass X-ray binary CXOGC J174540.0-290031 \citep{Porquet+2005}.

\begin{figure*}
\label{fig:fig1}
\begin{center}
\hbox{
\hspace{-0.cm}
\includegraphics[width=0.3\linewidth]{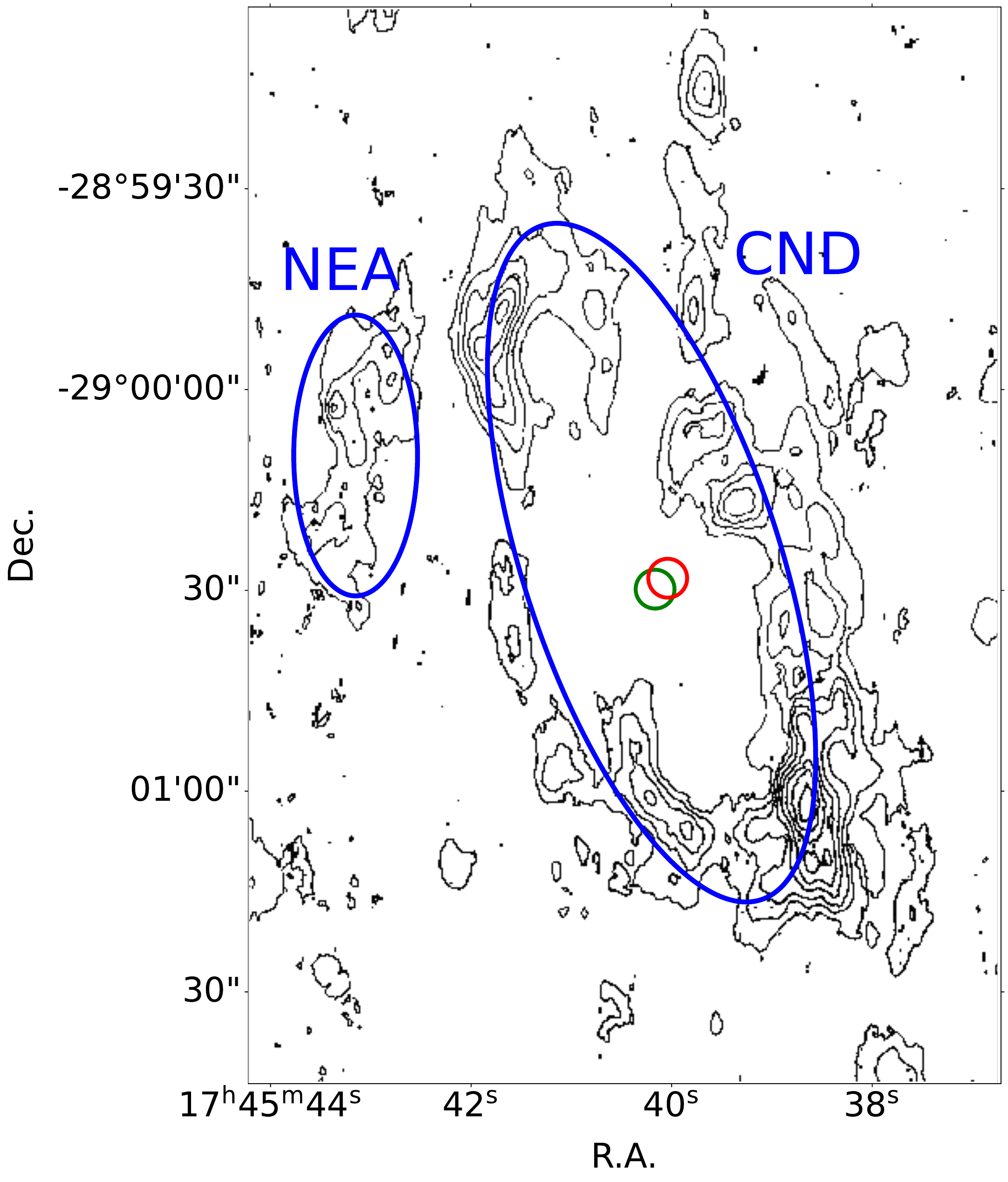}
\hspace{0.cm}
\includegraphics[width=0.38\linewidth]{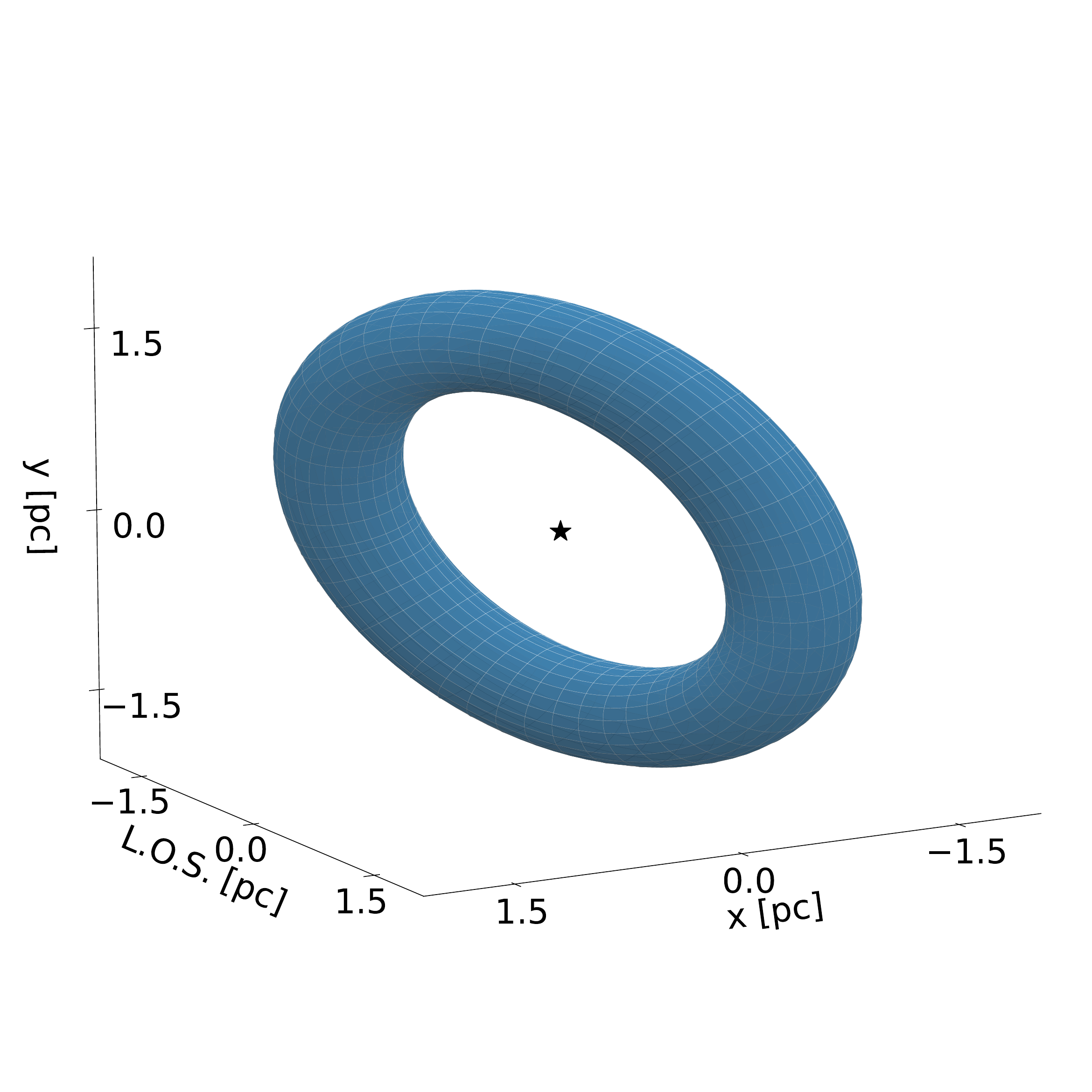}
\hspace{-1.5cm}
\includegraphics[width=0.38\linewidth]{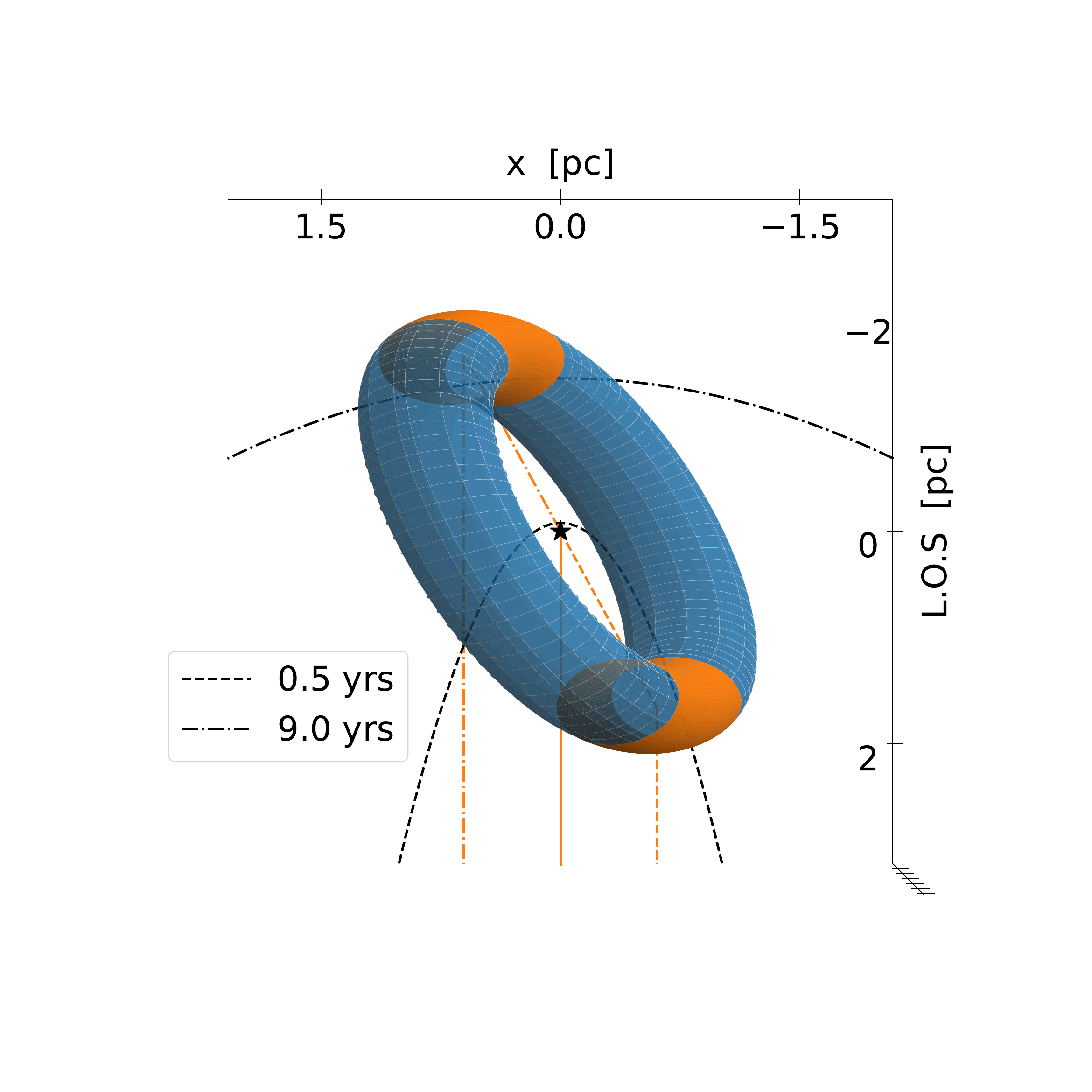}
}
\end{center}
\caption{\footnotesize Observation and schematic view of the CND.
{\itshape Left panel}: HCN contours of the CND as presented in \citet{Cristopher+2005}. The two ellipses highlight the proper CND and the NEA. The red circle marks the position of \sgra and the green one the position of the magnetar SGR J1745-2900. {\itshape Middle panel}: A 3D model of the CND. ($x,y$) is the plane of the sky (R.A., Dec.). The torus has an inclination of $\simeq 70^\circ $ with respect to the line of sight (LOS), and $\simeq 20^\circ $ with respect to the $y$-axis to match the observed position angle. The star marks the position of \sgra. {\itshape Right panel:} The ring is projected in the $x-$LOS plane (top view). For a transient close to the position of \sgra, such as SGR J1745-2900, the two orange highlighted parts of the CND are the first and the last to produce a fluorescence signal. The iso-delay paraboloids  \citet{Sunyaev+1998} are plotted as dashed (six-month delay) and dashed-dotted (nine-year delay) lines, respectively. The time delay measured by the observer is due to the different light paths reported in the same scheme according to the line style used for the parabolas. The solid line is the light path from the transient to the observer.}
\label{fig:CND}
\end{figure*}

The molecular clouds closest to \sgra \ are located in the circumnuclear disk (CND), a torus-like structure surrounding \sgra. The inner and outer radii of the CND are, respectively, $\simeq 1.5$ pc and $\simeq 2.0$ pc \citep{Lau+2013}, but clouds related to the CND can be found up to $\simeq 7$ pc from \sgra \citep{Genzel1989}. The inclination angle of the CND is estimated to be $\simeq 70^\circ$, its western side presumably tilted toward the Earth \citep{Lau+2013}. Dynamically, the CND shows quasi-Keplerian rotation around the SMBH, with a velocity of $\simeq 110$ km s$^{-1}$. The total mass is estimated to be between $10^4$ and $10^6$ \Msun, whereas the hydrogen column density is between $10^{22}$ and $10^{24}$ cm$^{-2}$ \citep{Wright+2001, Hsieh+2017,Hsieh+2021,Lau+2013,Martin+2012,Torres+2012,Zhao+2016,Tsuboi+2018,Dinh+2021}. \citet{Mossoux+2018} found the first evidence of the CND footprint in the X-ray band, seen as a depression in the X-ray luminosity around \sgra. Fig.\ref{fig:CND} (left panel) displays the HCN (hydrogen cyanide) contours of the CND as presented in \citet{Cristopher+2005}. Together with the CND, the so-called northeastern arm (NEA) is also indicated. The middle panel shows a 3D model of the CND as reconstructed from observations. 

By studying the reprocessed light from the CND we can get a better description of both the CND and the illuminating source. Indeed, the fluorescence intensity is directly related to the source luminosity and distance, and to the cloud optical depth. The geometry of the molecular material and the relative positions of clouds and sources also affect the observed signal and its time evolution \citep{Sunyaev+1998}. 
The right panel of Fig.~\ref{fig:CND} shows how the radiation front stemming from a source supposedly located at the center of the system scans the CND. An observer would first detect the radiation propagating directly from the source, then after a delay of 0.5 years the fluorescence from the near side of the CND, and finally after 9 years the reflection from the far side of the CND. Iso-delay curves are paraboloids and their focus coincides with the central source \citep{Kapteyn1902,Couderc1939,Sunyaev+1998}.

Our work aims to constrain the properties of the CND through a spatial and  spectral analysis of past X-ray observations of the GC. X-ray transients occurring in the outskirts of \sgra \ are very well monitored, as the GC is routinely  observed\footnote{\url{www.swift-sgra.com}} with short ($\sim$ 1 ks) exposures by Swift \citep{Degenaar+2015} almost daily. This fact, along with the large number of longer observations ($\gtrsim 10$ ks) accumulated in the last $\simeq 20$ years by XMM-Newton and Chandra, enable us to investigate the presence of any reflected light from the brightest transients. XMM-Newton, thanks to its large effective area at 6.4 keV and good PSF, is particularly well suited to perform this type of study. An illuminating source candidate of particular interest that we focus on is the magnetar SGR 1745-2900 that went into outburst in late April 2013 \citep{Rea+2013,Mori+2013}. The echo of the outburst is expected to be detectable as  iron fluorescence since the magnetar is orbiting close to \sgra, and hence we can predict the geometry and the time at which the signal should be visible. Moreover, as the magnetar outburst lasted for more than one year, X-ray observations of the GC with XMM-Newton (typically a few days per year) are likely to detect the fluorescence signal.

The paper is structured as follows. Section~\ref{sec:data} presents the dataset used, while Section~\ref{sec:calibration} focuses on the spectral analysis required to have a precise energy scale calibration. 
In Section~\ref{sec:variability} we study the CND variability in the Fe \Ka line band, using images and spectra. Candidate sources that could be responsible for the line variability are described in Section~\ref{sec:candidates}, while 
a discussion of our results, in particular the implications for the
physical structure of the CND, is reported in Section~\ref{sec:discussion}. Finally, Section~\ref{sec:conclusions} is devoted to concluding remarks.

\section{Data reduction}
\label{sec:data}
Given the high effective area at 6.4 keV of the EPIC-pn  onboard XMM-Newton, we considered observations taken with this instrument pointed at \sgra \ in 2000-2022. We excluded datasets in which bright transients significantly contaminate the emission in the CND region. In particular, the known X-ray binary AX J1745.6-2901 \citep{Muno+2003,Baganoff+2003} prevented us from analyzing data collected in the years 2007, 2008, 2013, 2015, and 2016.

We reprocessed the raw data with the XMM-Newton Science Analysis System (SAS 20.0.0) using the last release (as of July 1, 2022) of the valid current calibration files (CCF). The raw data were reprocessed with the task \verb|epproc|. The events were then filtered for flaring particle background. For this purpose, we selected periods of high background flaring activity by looking at the light curves of single events (\verb|"PATTERN==0"|) with energy in the range 10-12 keV. The good time intervals (GTIs) are defined as the periods in which the count rate is below 0.5 counts s$^{-1}$ in the whole CCD.
Since the CND is close to \sgra \, we removed periods in which flares from \sgra \ are present. To identify these periods we extracted light curves from a circular region of 12" radius. We then performed a Bayesian blocks decomposition of each light curve, as done in \citet{Ponti+2015}. 
We restricted the analysis to observations with a scientific exposure (after filtering) longer than 8 ks, in order to have enough statistics in the spectral calibration analysis. Since \sgra \ can be observed in early spring and early fall, we stacked observations belonging to the same season for each year, obtaining 12 data sets for a total of 44 accumulated spectra. We discarded sets with less than 50 ks exposure. Tab.\ref{table:1} lists a summary of the observations considered in this paper.

\section{Spectral analysis and calibration}
\label{sec:calibration}
\begin{figure*}
    \centering
    \includegraphics[width=0.8\linewidth]{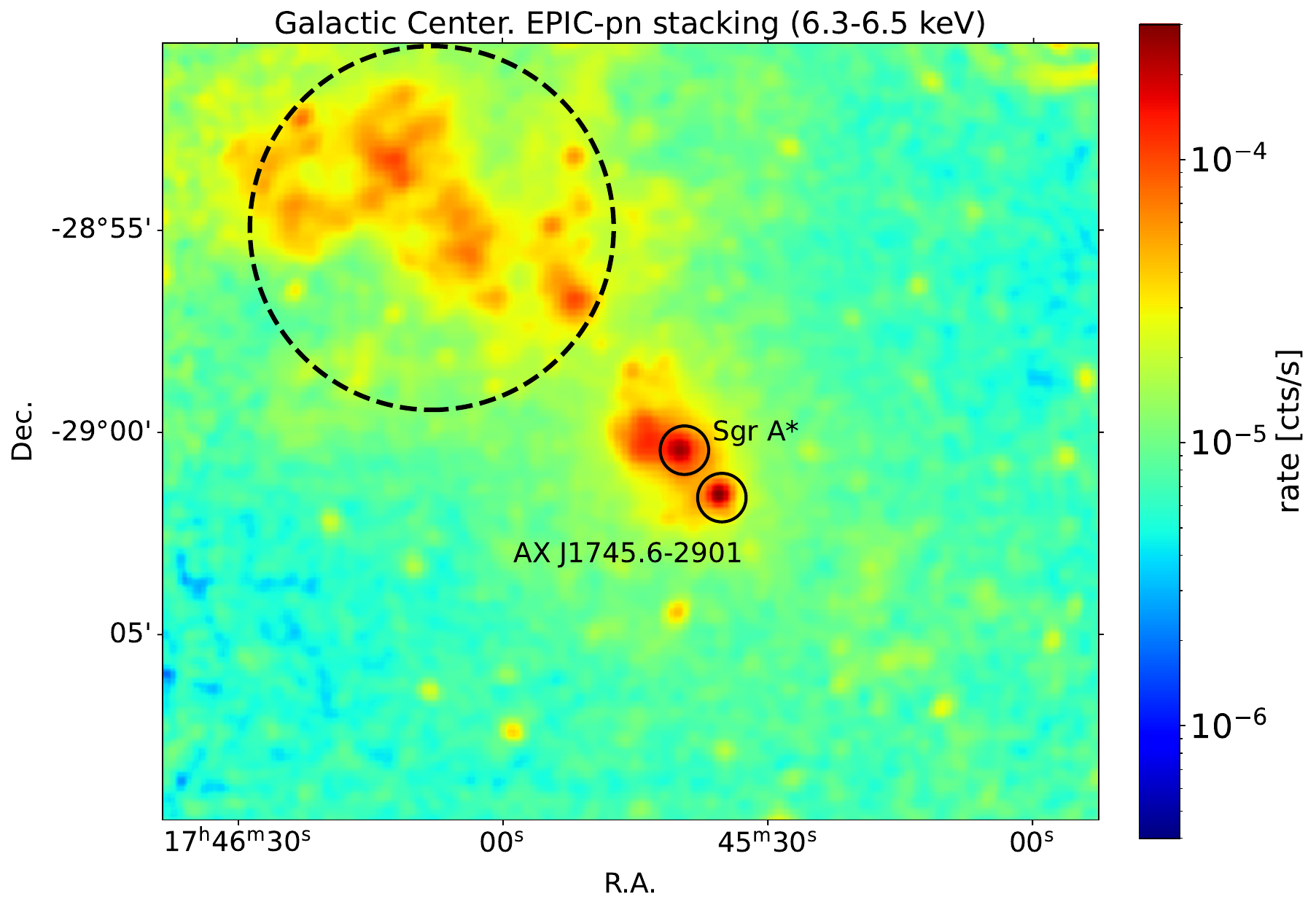}
    \caption{Galactic center as seen by XMM-Newton (EPIC-pn) in the 6.3-6.5 keV band (stacking of observations considered in this article). The dashed circular region marks the area used to accumulate the spectra for the energy calibration step. In that region, known massive GMCs reflect the past emission most probably associated with high-activity periods of \sgra . The extended emission region encompassing \sgra \ is the SN remnant Sgr A East. We also highlighted the position of the known transient AX J1745.6-2901, which highly contaminates the flux from the CND region.} 
    \label{fig:galactic_center}
\end{figure*}

Since the CND lies in the dense and rich environment of the GC, any flux variation of the CND occurs in a background-dominated regime.
As an example, the supernova remnant (SNR) Sgr A East \citep{Maeda+2002} accounts for most of the soft X-ray emission in the CND region (see Fig.\ref{fig:galactic_center}).
Though the SNR emission is not expected to have varied in the last 20 years, the GC shows a strong Fe XXV complex. This feature is a triplet that cannot be resolved by the EPIC camera aboard XMM-Newton, the centroid of which lies at $\simeq 6.65$ keV. It dominates the hard X-ray spectrum and contaminates any emission at 6.4 keV. Consequently, it was necessary to consider the energy uncertainty of the EPIC-pn camera on board XMM-Newton (about 12 eV). For this purpose, we characterized the energy of the fluorescence line in each observation through a detailed spectral analysis. In this way we aimed to minimize the contamination from the tail of the Fe XXV line to the fluorescence signal at 6.4 keV.

We first determined the actual energy of the Fe \Ka line by computing the spectrum on a large area where other emitting clouds are located. 
To this aim, we collected the spectrum on a circular area with a radius of 4.5 arcmin, centered on the position (R.A., Dec.) = (266.5333463,-28.9158656) deg. The extraction region is shown in Fig.\ref{fig:galactic_center} (dashed line). We refer to it as the outer region to differentiate it from the molecular clouds in the inner parsecs in the CND. We then proceeded to calibrate the spectral information\cprotect\footnote{ The events are selected with the expression \verb|"(FLAG==0) && (PATTERN<=4)"|. 
We remove out-of-time (OoT) events from the spectra \citep{XMM_manual}. 
RMF and ARF files were generated by running the \verb|rmfgen| and \verb|arfgen| commands, respectively, with the option \verb|extendedsource=yes| and \verb|detmaptype=flat|. }.

The spectral analysis was performed with PyXspec (based on Xspec version 12.12.1, \citet{Arnaud1996}), assuming a Poissonian distribution for the
events (\verb|cstat| statistic). We fit the spectra considering two energy ranges: one including the Fe lines complex (5.5-7.2 keV), and one (7.6-8.5 keV) centered on the instrumental Cu \Ka line (8.038 keV) due to the internal `quiescent' background.
We modeled each spectrum as the sum of a continuum power-law component and four emission lines (\verb|powerlaw+gaussian+gaussian+gaussian+gaussian| in Xspec), corresponding to Fe \Ka (6.4 keV), Fe XXV ($\simeq$6.65 keV), Fe XXVI ($\simeq$6.96 keV), and Cu \Ka, respectively. We fit the 44 spectra, keeping the line widths as common parameters in all the datasets. On the other hand, the energies of the lines are free. 

\begin{figure}
    \centering
    \includegraphics[width=\linewidth]{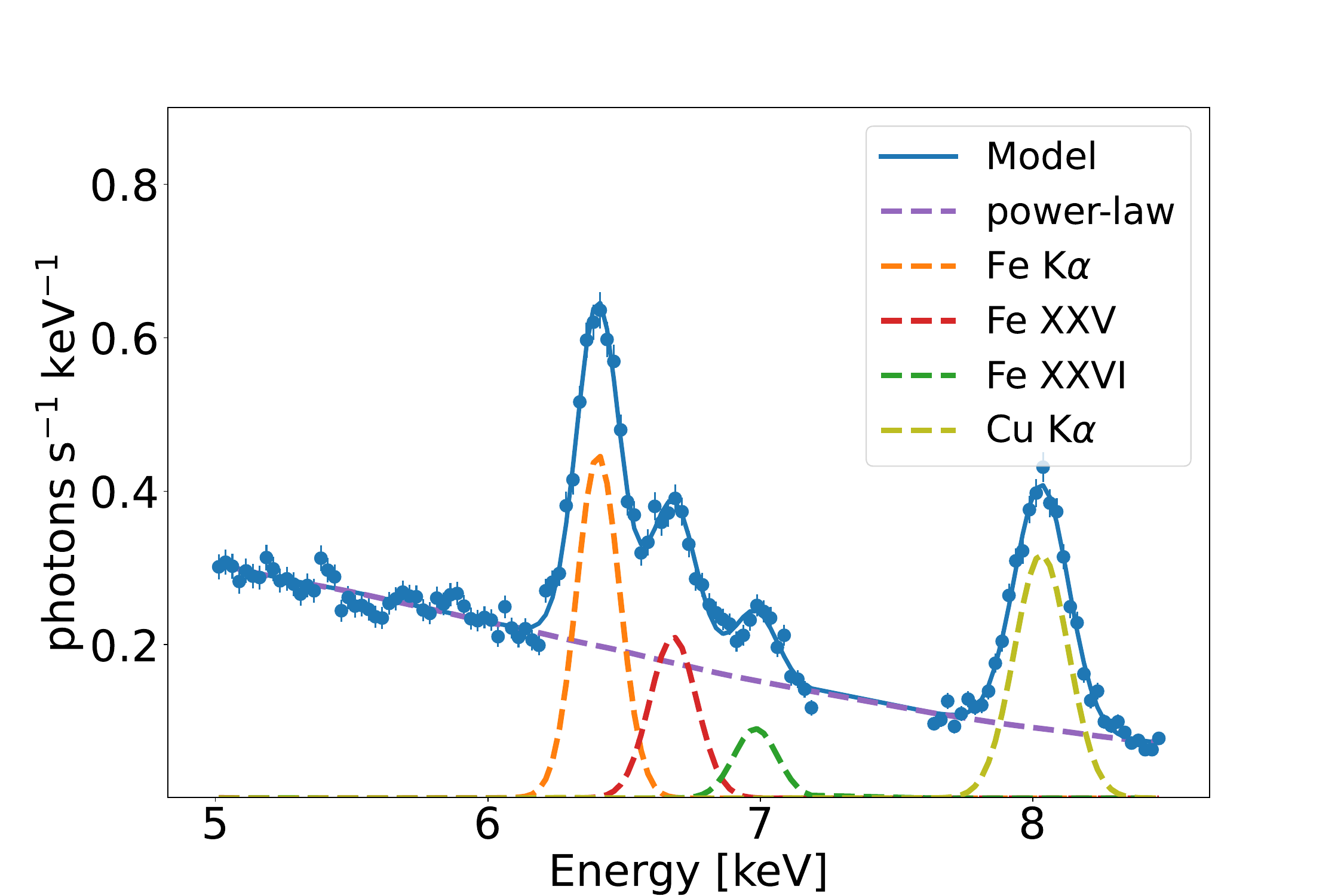}
    \caption{Calibration spectrum for observation no. 44 (ID 0893811301). The spectrum is accumulated in the circular dashed regions, enclosing giant molecular clouds, shown in Fig.\ref{fig:galactic_center}. We performed a fit in the 5.0-7.2 and in the 7.6-8.5 keV range. The fit model is a power-law continuum with four gaussian lines.}
    \label{fig:6_4_spectrum_fit}
\end{figure}

\begin{figure}
    \centering
    \includegraphics[width=\linewidth]{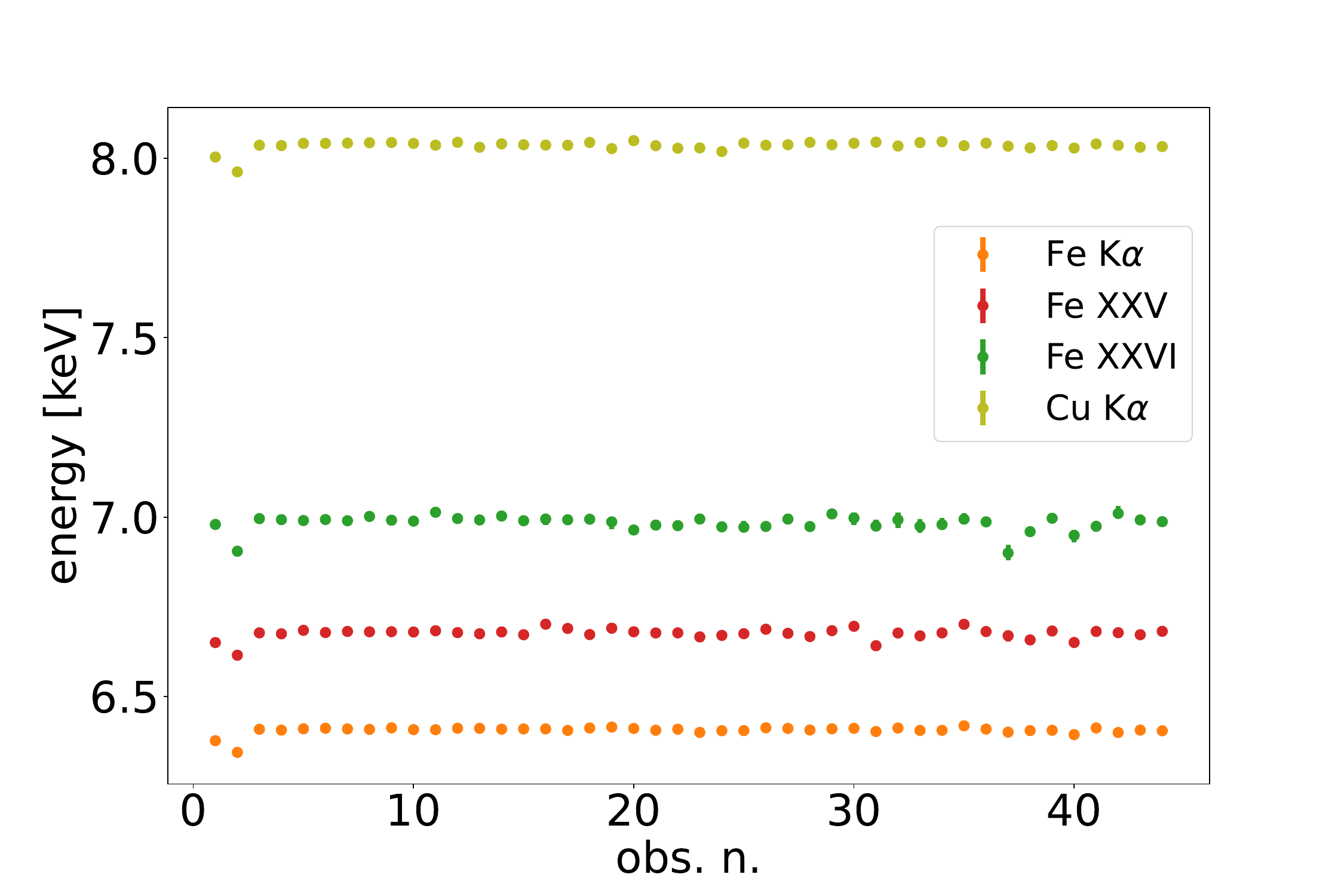}
    \caption{Energy of the  Fe \Ka (6.4 keV), Fe XXV (6.65 keV), Fe XXVI (6.96 keV), and Cu \Ka (8.038 keV) lines for the 44 observations considered in this work. The energy is the best-fit value, with an associated 1 $\sigma$ error.}
    \label{fig:6_4_centroids}
\end{figure}

\begin{figure}
    \centering
    \includegraphics[width=\linewidth]{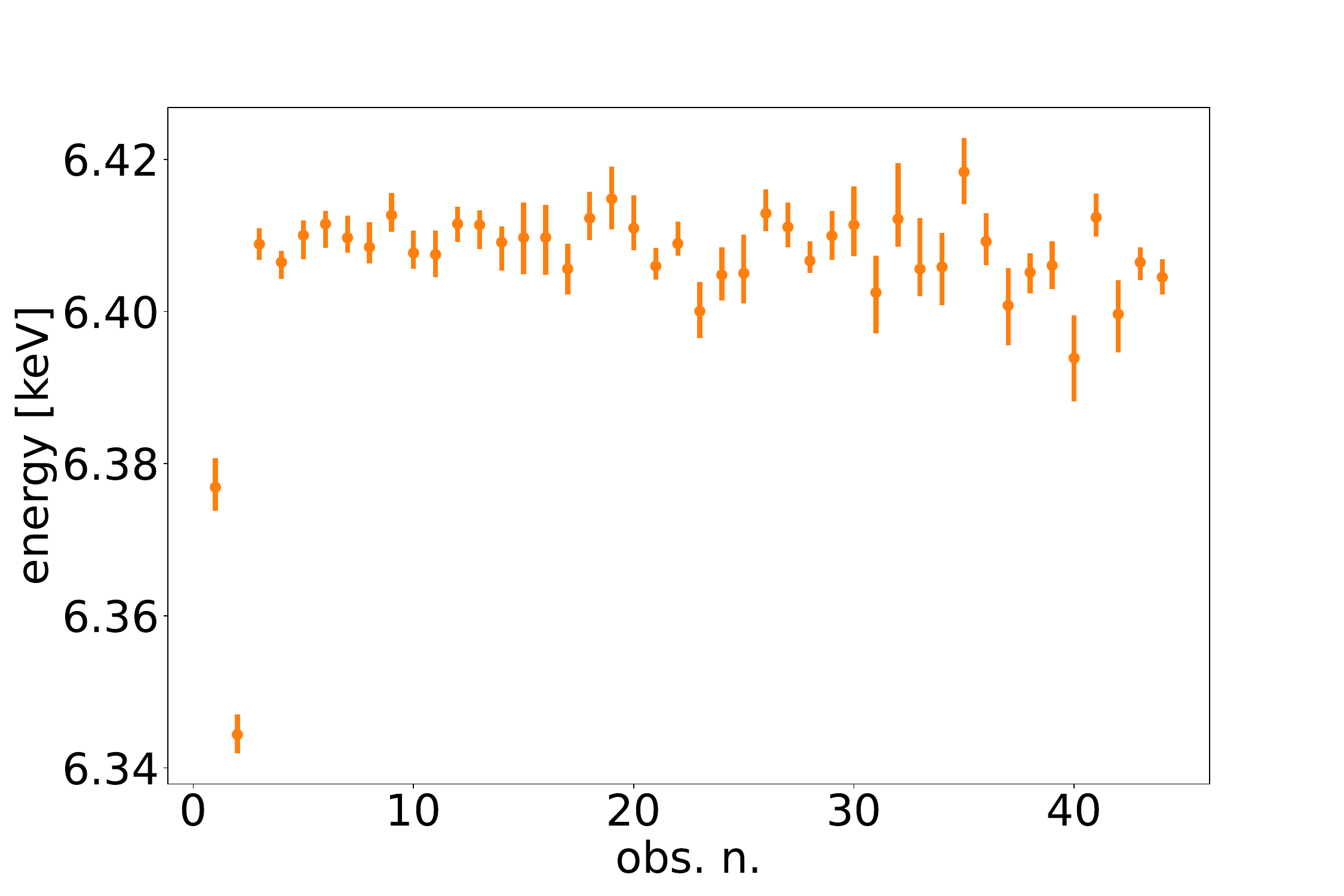}
    \caption{Energy of the Fe \Ka line for the 44 observations considered in this work, as in Fig.\ref{fig:6_4_centroids}. The energy measured for the first two observations highly deviates from the others, probably due to the fact that those were taken in extended full frame mode.}
    \label{fig:6_4_detail}
\end{figure}

As an example, in Fig.\ref{fig:6_4_spectrum_fit} we show data and the best-fit model for observation no. 44 (ID 0893811301, see Table~\ref{table:1}). The model provides a good fit to the data in all observations. The resulting C-statistic is less than 626 for 518 bins for all 44 spectra considered (a total reduced $\chi^2$ of 1.1). Fig.\ref{fig:6_4_centroids} shows the Fe \Ka, Fe XXV, Fe XXVI, and Cu \Ka line centroids along with their statistical uncertainties ($1 \sigma$), while  Fig.\ref{fig:6_4_detail} displays the Fe \Ka line energy only. Our analysis allowed us to constrain the energy of the Fe \Ka line with an uncertainty of about 10 eV and 2 eV for exposure of 10 ks and 50 ks, respectively. The first two observations present a larger deviation, common to all fitted lines. This is most probably associated with the full extended mode of XMM \citep{XMM_manual} used in acquiring this set. Consequently, we decided to discard the first two observations (set 1 in Tab.\ref{table:1}). In the next section, for each observation we will use as a reference the value we found here of the Fe \Ka line to generate images around 6.4 keV. Finally, we must note that in the analysis just described we did not take into account a spatial dependence of the energy scale calibration. Indeed, all observations considered in this work are pointed to \sgra, therefore the CND region lies in the central part of the CCD. On the other hand, the region that we used to accumulate the spectrum (circular dashed region in Fig.\ref{fig:galactic_center}) and perform the calibration step lies in the external part of the CCD. We return to this and discuss its implications in Section \ref{sec:variability}.

\section{Fe \Ka line variability}
\label{sec:variability}
\subsection{Imaging}
\label{sec:variability:imaging}
In order to investigate the fluorescence component, we created images (photon counts) and exposure-corrected images (count-rate) in the Fe \Ka line band for each seasonal set. We integrated photons on a bandwidth of 150 eV, between 6.3 keV and 6.45 keV\cprotect\footnote{The image pixel size is 4.35". Events were selected with the expression \verb|"#XMMEA_EP"|. 
For each observation we created a counts map and an exposure map, with the task \verb|eexpmap|. We combined them using the task \verb|emosaic|, to obtain counts, exposures, and count-rate images for each seasonal set.
The count-rate maps were convolved with a gaussian smoothing kernel (with $\sigma=2$ pixels).}. To check for Fe \Ka line emission and its variability, we subtracted count-rate images from a reference dataset, hence creating "rate difference" maps. We chose as our reference dataset that of spring 2011 (set number four in Tab.\ref{table:1}), it having the longest exposure and showing no evidence of bright transients. 

\begin{figure*}
    \centering
    \includegraphics[width=0.95\linewidth]{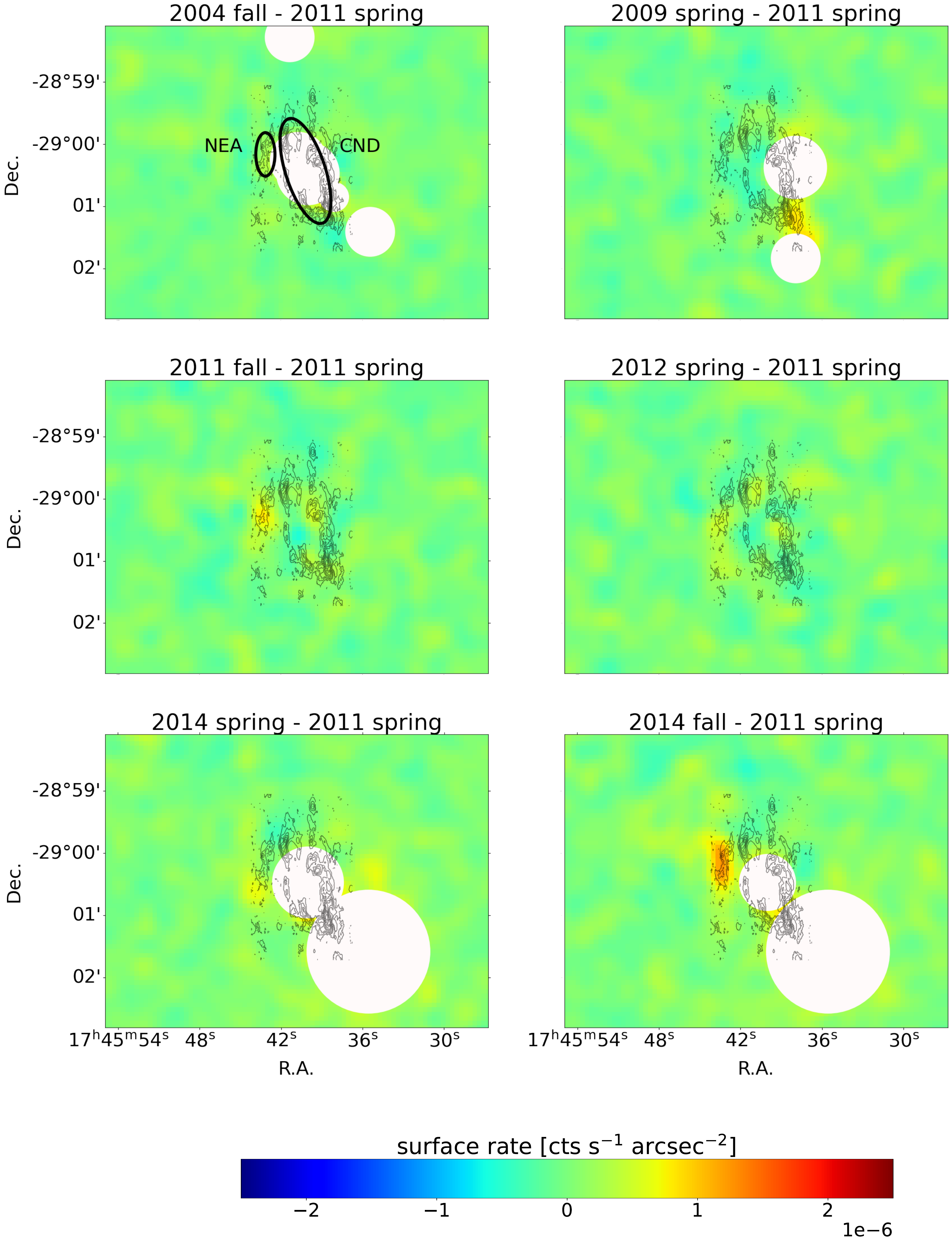}
    \caption{Difference between distinct datasets in the Fe \Ka line band. The colorbar ranges from -2.5 $\times 10^{-6}$ to 2.5 $\times 10^{-6}$ cts s$^{-1}$ arcsec $^{-2}$. In the upper-left panel, the large ellipse highlights the CND and the smaller one the NEA. Countours of the CND are taken from \citet{Cristopher+2005}. Transient point sources are masked. An increase in the Fe \Ka line flux is observed in the NEA during fall 2014. The bright transient of 2004 \citep{Porquet+2005} and the magnetar prevent us from studying the CND in 2004 and 2014.}
    \label{fig:differences1}
\end{figure*}

\begin{figure*}
    \centering
    \includegraphics[width=\linewidth]{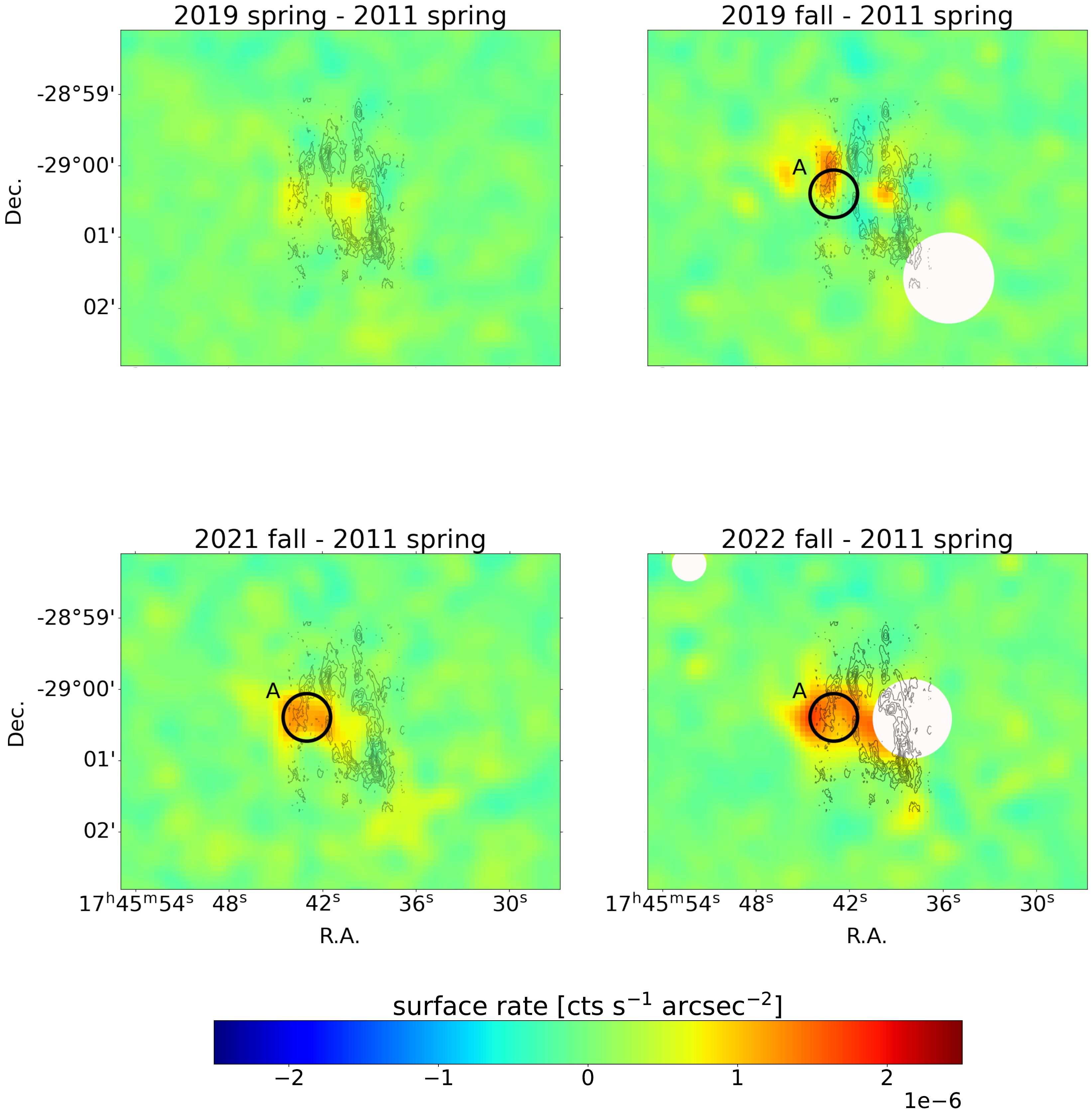}
    \caption{Difference between distinct datasets in the Fe \Ka line band. The colorbar ranges from -2.5 $\times 10^{-6}$ to 2.5 $\times 10^{-6}$ cts s$^{-1}$ arcsec$^{-2}$.  Transient point sources are masked. Fall 2019 and springs 2021-2022 show an increase in the count rate in the eastern part of the CND.}
    \label{fig:differences2}
\end{figure*}

Fig.\ref{fig:differences1} and \ref{fig:differences2} show such rate difference maps. The CND HCN contours \citep{Cristopher+2005} are superimposed. As the most significant variations are due to transient point-like sources, we selected and masked the brightest ones (except \sgra) with circular masks.  In the difference maps an excess in the count rate, in regions consistent with the CND position, is visible in falls 2014 and 2019 in the NEA. Another significant Fe \Ka line signal appeared in 2021 and increased through the 2022 images in the eastern part of the CND. The extended shape of the Fe \Ka excesses in the rate difference maps suggests that the signal cannot be associated with unresolved transient point-like sources. Moreover, the enhancement is not observed in the continuum.

It must be stressed that the rate difference images could be affected by uncertainties in the energy scale calibration. Indeed, the integration band used to generates images includes the tail of the brighter Fe XXV line (see Fig.\ref{fig:ka_fit} and Section\ref{sec:spectral_analysis}). Therefore, a slight shift in the line can result in an excess at about 6.4 keV. A shift is possible because of uncertainties in the energy scale calibration (via the charge transfer inefficiency (CTI) of the CCD). We tried to mitigate this effect by looking at the energy of the Fe \Ka line in other molecular clouds in the field of view. However, the CTI correction is spatially dependent across the CCD. In all the observations considered, the CND lies in the central part of the CCD, while the outer region used to verified the position of the Fe \Ka line is in the outer part of the CCD. Therefore, the images are still affected by this systematic effect. In the next section we perform a spectral analysis to remove this limitation.

\subsection{Spectral analysis}
\label{sec:spectral_analysis}

\begin{figure}
\centering
    \centering
    \includegraphics[width=1.0\linewidth]{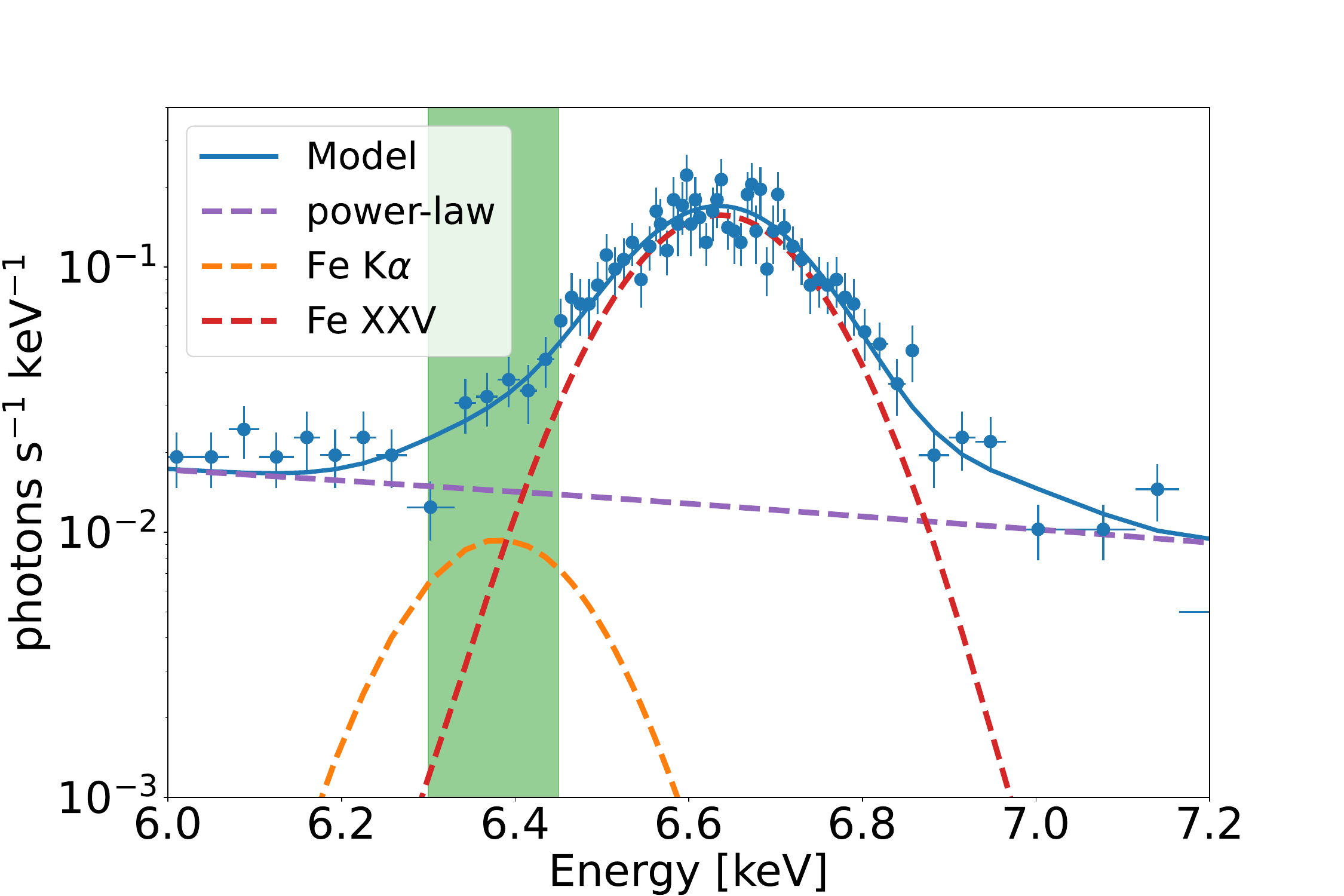}
\caption{Spectrum for observation no. 44 (ID 0893811301, spring 2022) in region A. We performed a fit in the 5.0-8.0 keV energy band. The fit model is a power-law continuum with two narrow gaussian lines. The Fe \Ka line energy (6.4 keV) was kept at a fixed distance (-0.25 keV) from the Fe XXV line. The green area highlights the energy range used to generate the images analyzed in Section \ref{sec:variability}.}
\label{fig:ka_fit}
\end{figure}

In order to obtain a better characterization of the Fe \Ka signal and quantify the significance of the variations once the uncertainties in the energy calibration were taken into account, we performed a detailed spectral analysis. In this way the calibration issue of the imaging thread could be treated properly. We chose a circular region (region A in the images) that encloses the eastern region and partially the NEA, in order to describe the variability in the Fe \Ka line over the years. For each seasonal set, we considered the observation with the longest exposure. 
In the CND region, the Fe XXV line is the most prominent spectral feature. The Fe \Ka line provides only a smaller contribution, as shown in  Fig.\ref{fig:ka_fit}, which displays the spectrum accumulated in region A during spring 2022 (cfr. Fig.\ref{fig:6_4_spectrum_fit}).

\begin{figure}
\centering
    \centering
    \includegraphics[width=1.0\linewidth]{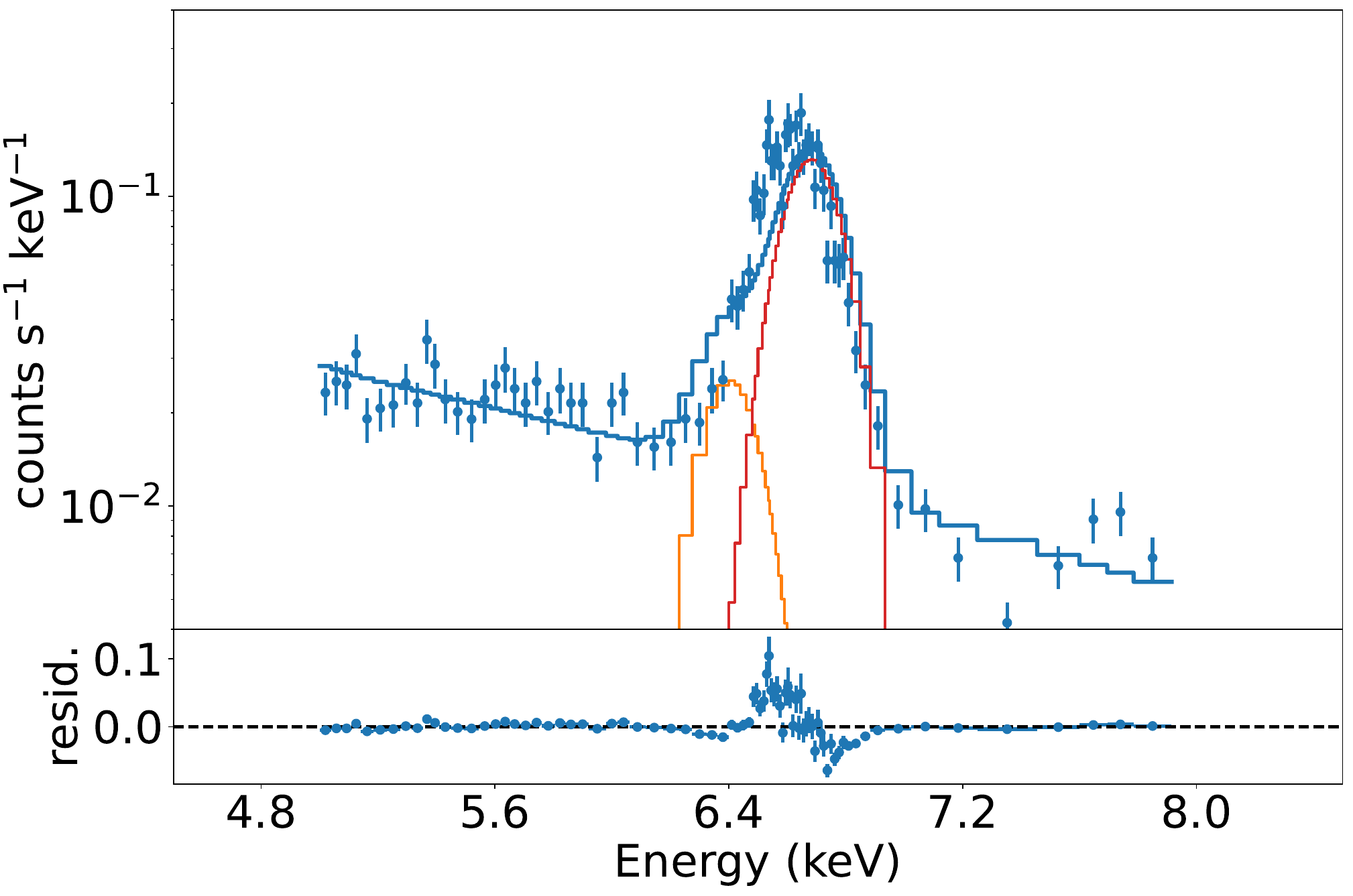}
\caption{Spectral fit for observation no. 44 (ID 0893811301, spring 2022) in region A. We performed a fit in the 5.0-8.0 keV energy band. The fit model is a power-law continuum with two lines. The energy and width of each line was fixed to the value found in the calibration step (Section \ref{sec:calibration}). The bottom panel shows the residuals (data-model).}
\label{fig:fit1}
\end{figure}

\begin{figure}
\centering
    \centering
    \includegraphics[width=1.0\linewidth]{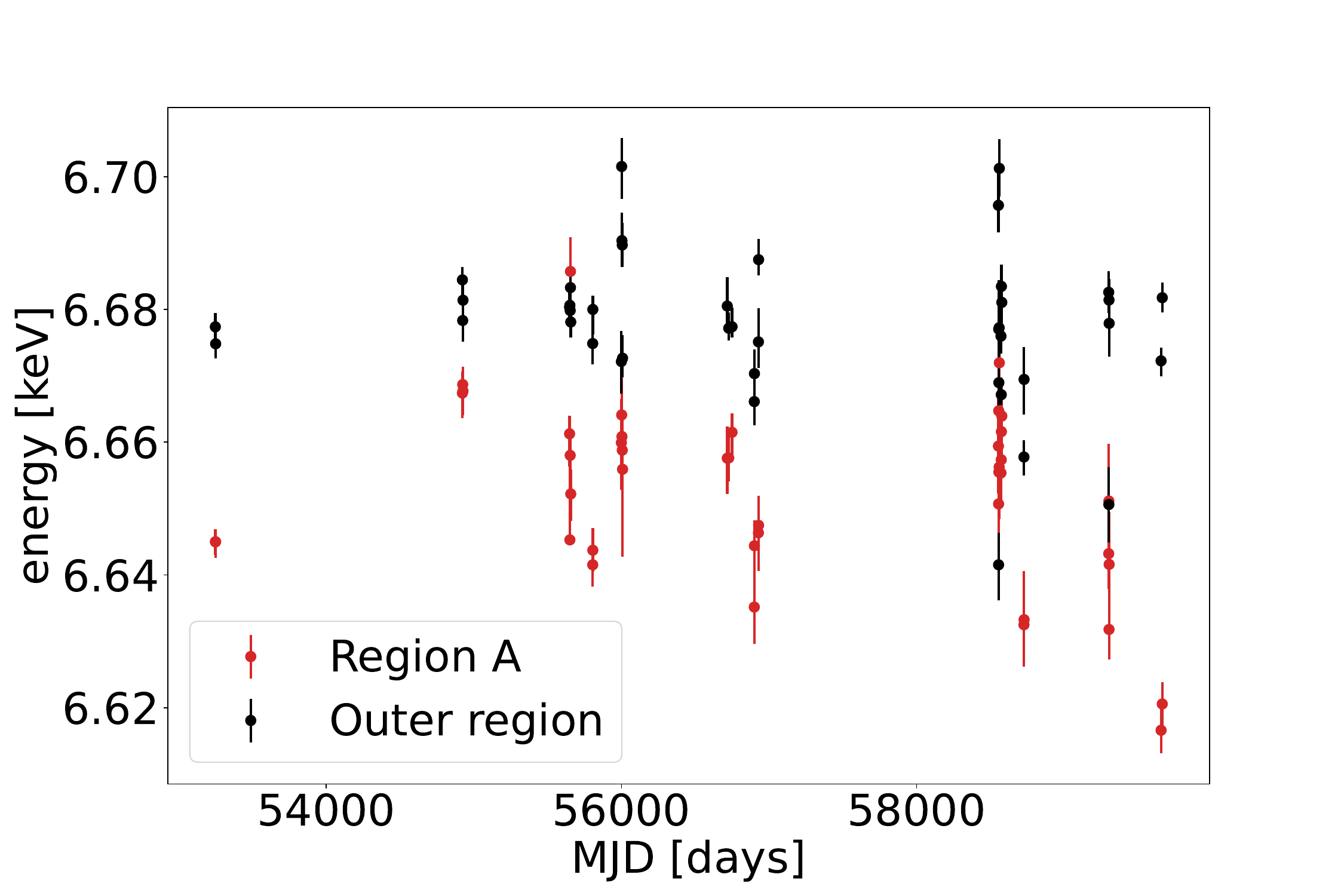}
\caption{Best-fit value of the energy of the Fe XXV line in region A for the observations considered in this work (red data points). The energy is the best-fit value obtained for a power-law and gaussian line model. The error bars refer to the associated 68\% confidence interval. MJD refers to the start of the observation date. The black data points are the centroids of the Fe XXV triplet measured in the outer region and displayed in Fig.\ref{fig:6_4_centroids}.}
\label{fig:fexxv}
\end{figure}

\begin{figure}
\centering
    \centering
    \includegraphics[width=1.0\linewidth]{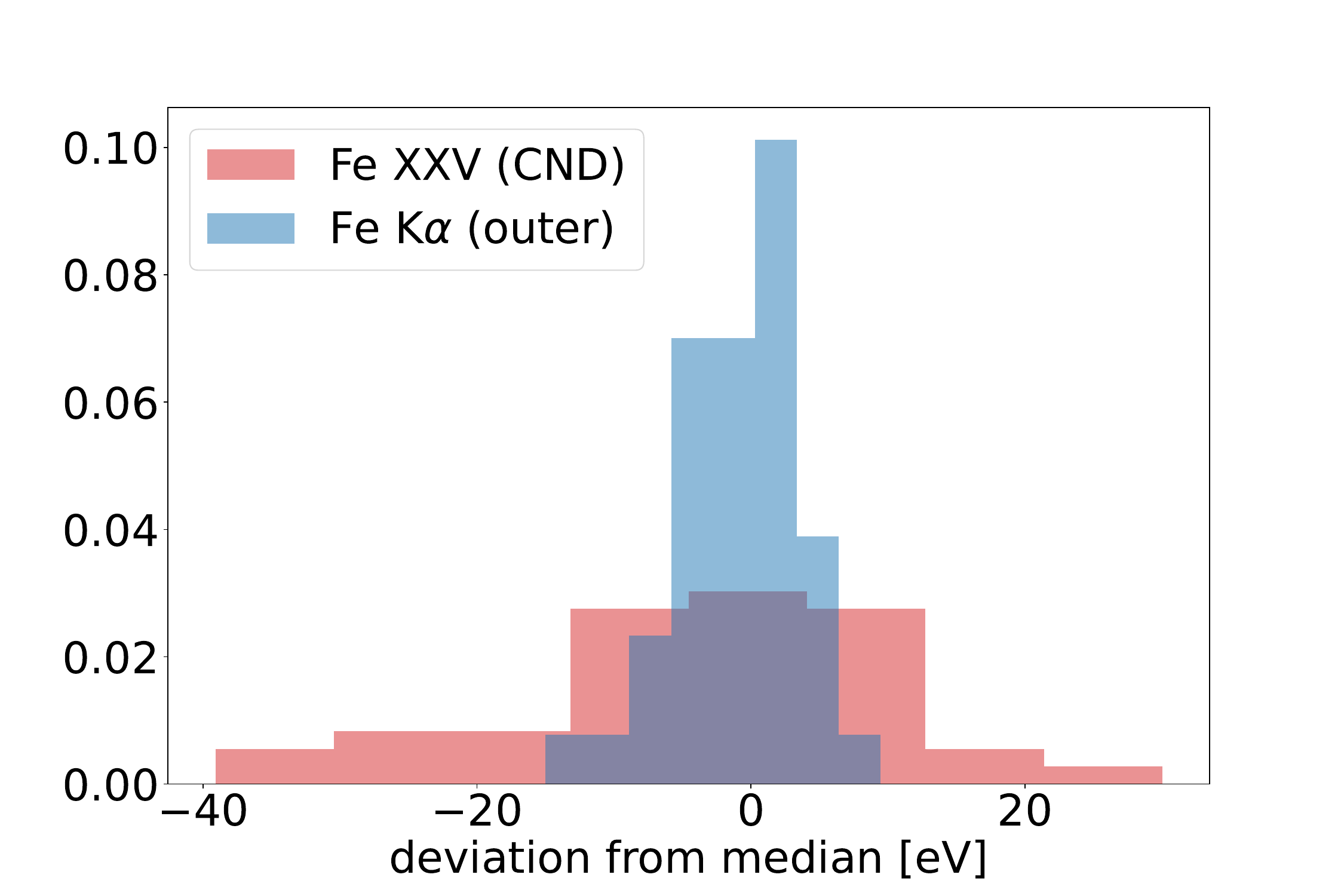}
\caption{Normalized histogram of the centroids of the Fe XXV line energy in the CND (region A), and the Fe \Ka (in the outer region), after the subtraction of the median value.}
\label{fig:hist_ka_xxv}
\end{figure}

We first attempted to fit the spectra in the 5.0-8.0 keV energy band, with a continuum component (powerlaw) with two gaussian lines for the Fe \Ka and Fe XXV. We fixed the energy and the width of the two gaussians at the values found in Section \ref{sec:calibration} for each observation. The model failed to effectively reproduce the iron complex. Adding a third Fe XXVI component did not improve the fit. Fig.\ref{fig:fit1} displays the spectrum and best fit for observation no. 44 (spring 2022) in region A. The fit shows large residuals in the iron complex. To verify this behavior, we fitted the spectra of all observations with a simple power-law and gaussian line model, keeping all the parameters free and independent. Fig.\ref{fig:fexxv} displays the position of the Fe XXV energy as a function of the observation date (red points). In the same plot the distribution of the Fe XXV centroid in the outer region (black points) is reported. The energy of the Fe XXV is systematically lower in the CND region. This is due to the different properties of the plasma that translates in a different centroid of the Fe XXV triplet. Indeed, one originates from the Galactic ridge X-ray emission and the other from the SNR Sgr A East \citep{Maeda+2002}. The distribution in the CND (red points) presents a larger scatter but is still comparable to what is observed in the outer region (black points), analyzed in Section \ref{sec:calibration}. Moreover, the scatter is much larger in comparison to the Fe \Ka distribution in the outer region. This behavior is shown in Fig.\ref{fig:hist_ka_xxv}, in which we plot the histogram of the energy distribution of the Fe XXV in the region A and the Fe \Ka in the outer region. The large scatter of the Fe XXV triplet affects the imaging analysis. For example, the energy of the line is lower in the falls of 2014 and 2019 and in spring 2021 (see Fig.\ref{fig:fexxv}), reaching $\simeq 6.62$ keV in the last set (spring 2022). On the other hand, the energy of the 6.4 keV line that we used to select the energy band to generate the images does not show this large variation. Therefore, when creating the image for some of the seasonal sets a larger fraction of the Fe XXV line's tail was included in the energy band, resulting in an higher count-rate in the map. At the moment we cannot claim whether the energy shift is due to any physical phenomena or, rather, reflects some calibration issues.

To take into account the variability of the Fe XXV line and the Fe \Ka one we therefore performed a new fit, allowing the energy scale to shift. We chose a fixed distance (0.25 keV) between the Fe \Ka line and the Fe XXV line and left the energy of the Fe XXV as a free parameter. The 0.25 keV value is the distance between the median energy of the Fe \Ka line measured in the outer molecular clouds (Fig.\ref{fig:6_4_centroids}) and the median energy of the Fe XXV centroid measured in region A. We also estimated the width of the Fe XXV directly from the spectra computed in region A.
Since the bulk of the emission in the considered region is due to the SNR Sgr A East, the shape of the line is not expected to vary in a timescale of years. Then, in order to determine the width of the line, we simultaneously fit the unbinned spectra in the 5.0-8.0 keV range with a simple power-law and Gaussian line model. We excluded sets from 2019, where a Fe \Ka signal is present. The best-fit value for the line width is $(27 \pm 5)$\,eV. 
With this value to hand, we performed a second fit, now adding a narrow Fe \Ka line (0-width \verb|gaussian| component in Xspec).
We performed a Bayesian parameter estimation using BXA with PyXspec \citep{Buchner+2014}. For each fit, there are six free parameters (two for the power-law, three for the Fe XXV line, and the normalization of the Fe \Ka line, but not its energy, since it is tied to the Fe XXV one). We adopted a uniform prior for the normalization of the lines and the energy of the Fe XXV. A gaussian prior for the width of the line was taken from the previous fit ($(27 \pm 5)$\,eV). 

\begin{figure}
\centering
    \centering
    \includegraphics[width=\linewidth]{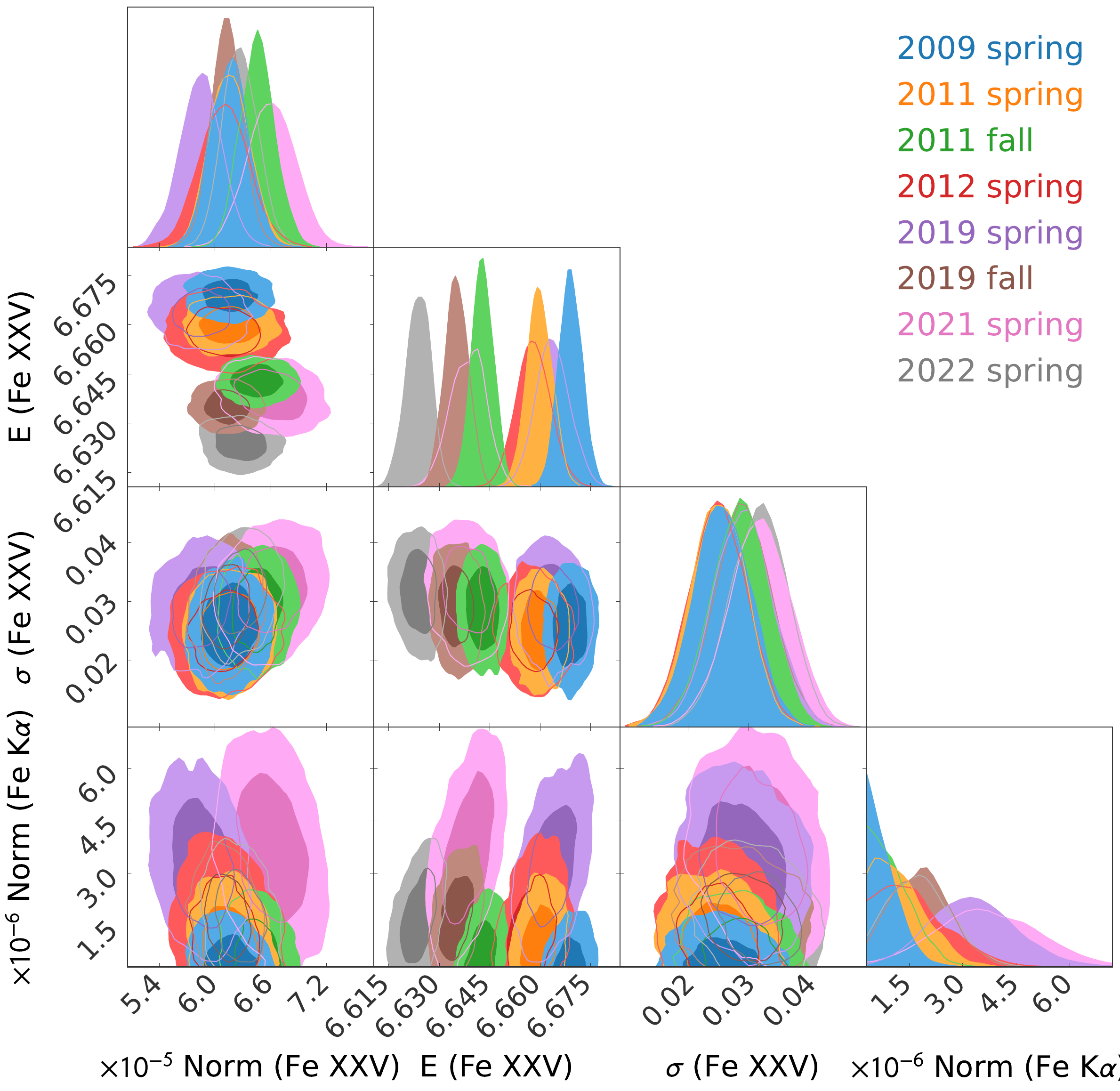}
\caption{Corner plot for the marginal parameter distributions in region A. The parameters in the x-axis are the normalization (photons s$^{-1}$ cm$^{-2}$) and the energy (keV) of the Fe XXV line, the width of the Fe XXV line, and the normalization of the Fe \Ka line. The graph was made with {\itshape pygtc} \citep{Bocquet2016}.}
\label{fig:corner}
\end{figure}

\begin{figure}
\centering
    \centering
    \includegraphics[width=\linewidth]{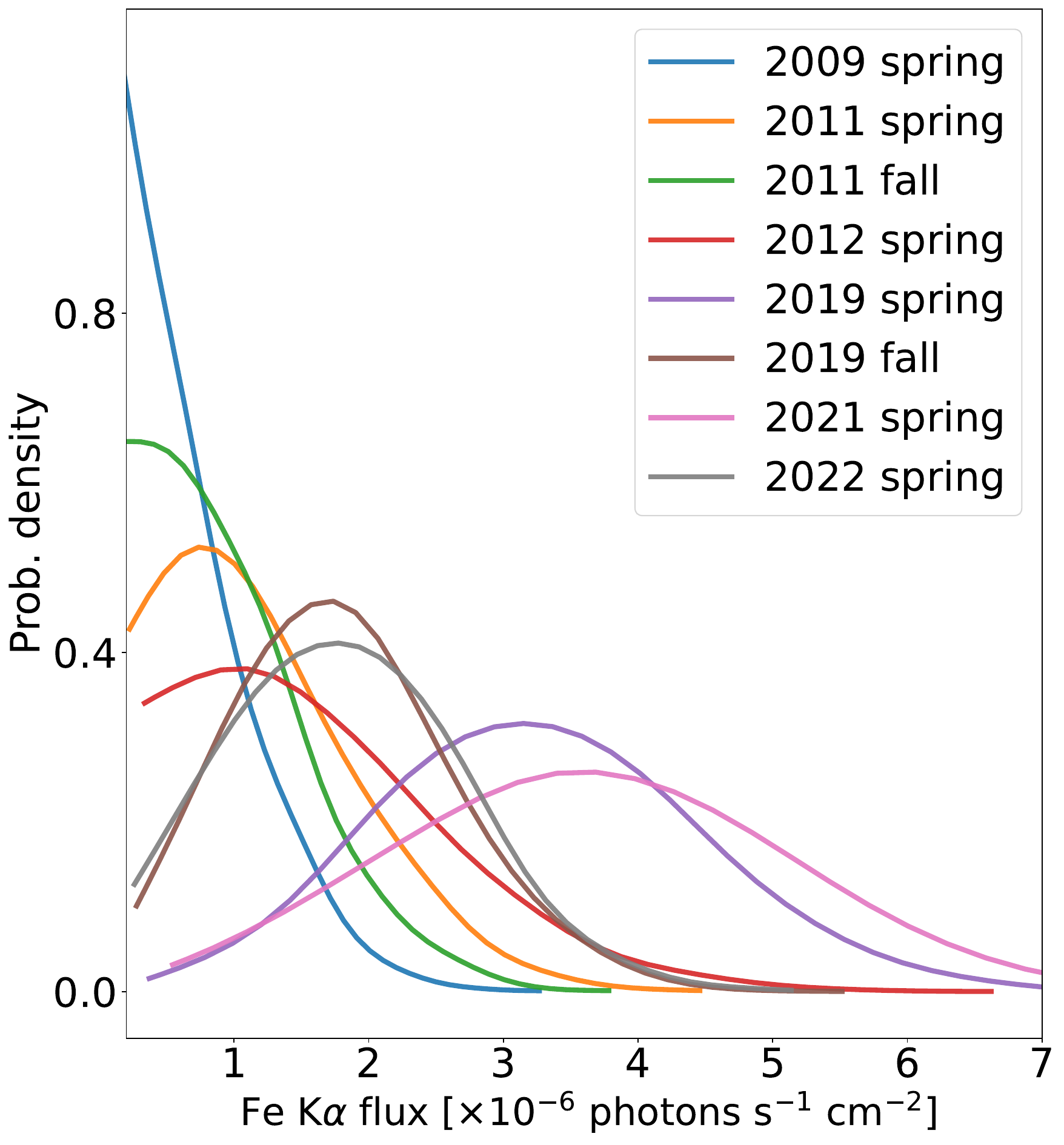}
\caption{Posterior distribution of the Fe \Ka flux in region A (units of photons s$^{-1}$ cm$^{-2}$). The spectrum is computed for the longest observation in each seasonal set. 
The energy of the line is tied to the energy of the Fe XXV. We excluded sets in which bright transients contaminate region A (2004, 2014).}
\label{fig:posterios_shift}
\end{figure}

The corner plot in Fig.\ref{fig:corner} shows the behavior of the various parameters for region A, displaying once again the shift toward lower energy of the Fe XXV line in the last years, possibly associated with a nonoptimal calibration of the energy scale. 
Fig.\ref{fig:posterios_shift} shows the posterior distributions of the  Fe \Ka flux (or equivalently, in Xspec, the line normalization). Before 2019 the signal is consistent with a null normalization. Starting from 2019, a positive flux is instead preferred by the fit procedure in region A. 
Therefore, the general behavior of the images is reflected in the Fe \Ka line flux in the spectra.

From these posterior distributions we computed the confidence intervals of the Fe \Ka flux. In the chosen region, the 95\% confidence interval is not consistent with a null flux in spring 2019 and spring 2021. The 68\% confidence interval is $[2.0, 4.3 ]\times 10^{-6}$ and $[2.2, 5.0 ]\times 10^{-6}$\,photons s$^{-1}$ cm$^{-2}$, respectively, for these two seasonal sets. For fall 2019 and spring 2022 the 68\% confidence interval is still consistent, but slightly lower: $[0.8, 2.4 ]\times 10^{-6}$ and $[0.9, 2.7 ]\times 10^{-6}$\,photons s$^{-1}$ cm$^{-2}$, respectively.
Combining the results from the four seasonal sets we derived an interval, $[0.8, 3.5 ]\times 10^{-6}$\,photons s$^{-1}$ cm$^{-2}$ (68 \% confidence interval), with a mode value of $2 \times 10^{-6}$ photons s$^{-1}$ cm$^{-2}$.

Finally, we verified that alternative priors for the line normalization (e.g., log-uniform, Jeffreys prior, \citet{Jeffreys+1946}) do not change the results. The increased intensity of the Fe \Ka line that is observed in the eastern region from 2019 can explain only marginally the large energy shift of the nearby 6.65\,keV line. The shift is not observed in the Fe \Ka line in the outer region, and it differs from the Fe XXV variability in the outer region even if large scatter is present.
For this reason it is likely due to spatially dependent calibration of the energy scale.

\section{Source candidates}
\label{sec:candidates}
Many transient sources lying close to the CND could be the source of primary radiation, then generate the fluorescence emission. Here we discuss which transients could be associated with the Fe \Ka excesses described in the previous section. For the sake of simplicity, we consider only those transients that have projected positions lying within the CND. For all sources we assume a distance of 8.3 kpc from Earth. 

\begin{itemize}
\item {\itshape \sgra:} The first and more obvious source is \sgra \ itself. Many works suggest that the emission from the MW supermassive black hole is the ultimate cause of the 6.4 keV fluoresce observed in the giant molecular clouds in the CMZ. \sgra \ presents a quiescent 2-10 keV luminosity $\simeq 2 \times 10^{33}$\, erg s$^{-1}$ \citep{Baganoff+2003}. X-ray flares are observed daily \citep{Baganoff+2001, Porquet+2003, Neilsen+2013, Ponti+2015} but even the brightest ones (with luminosities hundreds of times in excess of the quiescent level\citep{Haggard+2019}) are much too brief (lasting at most a few hours) to be responsible for any detectable fluorescence signal from the CND. 

\item {\itshape SGR J1745-2900:} A candidate of certain interest is the magnetar SGR 1745-2900, which was observed outbursting starting on April 24, 2013$^{\rm th}$ \citep{Degenaar+2013,Kennea+2013,Mori+2013,Rea+2013,Rea+2020}. The echo from such an outburst should be detectable for two reasons. First, the magnetar is located at a projected distance of just 0.097\,pc from \sgra, most probably orbiting \sgra \citep{Rea+2013,Bower+2015}. Secondly, the X-ray luminosity remained  $\gtrsim 10^{35}$ erg s$^{-1}$ for more than one year after the start of the outburst. The spectrum can be well described by a black body ($T \simeq 0.8$\,keV) with a high energy power-law tail \citep[$\propto E^{-\Gamma}$, with  $\Gamma \simeq 2.0$,][] {CotiZelati+2015, CotiZelati+2017}. The total energy emitted in the first half year of the burst in the [2-10]\,keV band was $\simeq 3.3 \times 10^{42}$\,erg while, in the first year, it reached $\simeq 5.2 \times 10^{42}$\,erg. The imprint of the magnetar on the CND emission also gives fundamental information about the position of the source. For example, if SGR J1745-2900 is orbiting close to \sgra (at, say, $0.1$\,pc), then the signal is expected to appear in different portions of the CND at different times, according to the geometry of the CND. If the CND is oriented so that its western part points at us, then the first echo light should appear $\simeq 6$ months after the outburst. The opposite region should be illuminated instead $\simeq 9$ years after the flaring event. 

\item {\itshape Swift J174540.7-290015:} On February 6, 2016, the Swift satellite detected $\simeq 16''$ north of \sgra \ a transient with an associated X-ray luminosity $\simeq 8.5\times 10^{35}$\,erg s$^{-1}$ \citep{Reynolds+2016} (assuming a distance of 8.3 kpc). The detection was performed during the first observational run of the season, and therefore the transient may well have been active before (from November to February), when the GC is not observable due to the solar constraint window. The transient remained bright for more than four months \citep{Mori+2019}. The luminosity continued to rise after the outburst, reaching a peak of $\simeq 1.15\times 10^{37}$\,erg/s. The source showed a typical spectral evolution of a low-mass X-ray binary (LMXB). Analysis of the first 33 days from the detection conducted by \citet{Ponti+2016} shows that the photon index increased from $\simeq 2$ to $\simeq 6$. 
In the first $\simeq 16$\,days (the "hard spectral state"), 
the spectrum can be described by a power law with photon index $\Gamma \sim 2.56$. The total emitted energy (2-10 keV) during this period is $\simeq 1.2 \times 10^{43}$\,erg. In the next $\sim 17$ days the spectrum can be well fit by an absorbed black-body radiation with $kT \sim 0.8$ keV. The total  emitted energy in this "soft spectral state" is $\simeq 1.2 \times 10^{43}$\,erg.
The transient faded by mid-March \citep{Mori+2019}. In July the luminosity was $< 10^{36}$ erg s$^{-1}$. In this last period, consistent with a typical evolution of LMXBs, the source reached a "low-hard state", with the spectrum that could be described by a power law component with $\Gamma \sim 1.8 \div 2.5$ \citep{Corrales+2017}. The luminosity remains $>10^{35}$ erg s$^{-1}$ from the beginning of April to at least the beginning of July.

\item {\itshape Swift J174540.2-290037:} The transient was discovered on May 28, 2016, located $\simeq 10''$ south of \sgra \citep{Degenaar+2016}, with [2-10]\,keV and a luminosity $\simeq 5 \times 10^{35}$\,erg s$^{-1}$. The luminosity increased to $\simeq 2.9 \times 10^{36}$\,erg s$^{-1}$ in June and fell to $\simeq 9.2 \times 10^{35}$ by the end of the month \citep{Degenaar+2016b}. The transient lasted for $\simeq 1$\,month, releasing a total [2-10]\,keV energy $\simeq 5 \times 10^{42}$\,erg \citep{Mori+2019}.

\end{itemize}

Regarding the two transients discovered by Swift, it should be noted that despite their small projected distance from \sgra, their real 3D positions are still unknown. This results in significant uncertainties when modeling such transients as sources illuminating the CND, as we discuss in the next session. 
The magnetar instead is expected to be responsible for a Fe \Ka fluorescence signal, since its position is close to \sgra.
Therefore, if one or both of these two transients are associated with the echo radiation, their contribution must be added to the one coming from the magnetar SGR J1745-2900. 
Tab.\ref{tab:candidates} lists the illuminating source candidates considered in this section along with their typical luminosity, duration, and total energy emitted. For \sgra\ we reported the typical mean luminosity and the typical duration of the flares. We reported also the total energy emitted in a year, which is derived from \citet{Ponti+2015}, who report a total fluence (flux times duration) of $1.4 \times 10^{-6} \ \text{erg cm}^{-2}$ in about 6.9 Ms of total exposure. Therefore, in a year we expected about $5 \times 10^{40}$ erg of total energy emitted by \sgra during its flaring state.

\begin{table}[htb]
    \centering
    \begin{tabular}{c|c|c|c}
    \hline \hline
    Source & luminosity & duration & energy \\
        & [erg s$^{-1}$] &  & [erg]\\
    \hline
     \sgra\ (flares) & $10^{34} \div 10^{36}$ & $\sim  \text{few} 10^{3}$ s & $ 5 \times 10^{40}$\\
     SGR J1745-2900  & $\sim 10^{35}$ & $\gtrsim$ 1 yrs & $5.2 \times 10^{42}$\\
     J174540.7-290015 & $\gtrsim 10^{36}$ & $\gtrsim$ 4 months & $2.4 \times 10^{43}$\\
     J174540.2-290037 & $\sim 10^{36}$ & $\gtrsim$ 1 month & $5 \times 10^{42}$ \\
    \hline
    \end{tabular}
    \caption{Summary of illuminating sources candidates for the fluorescence signal of the CND and their typical luminosity, duration, and energy released. For \sgra\ the average duration of the flare is a few thousand seconds. The energy is extrapolated for a year of flaring activity, assuming the fluence reported by \citet{Ponti+2015}.}
    \label{tab:candidates}
\end{table}

\section{Discussion}
\label{sec:discussion}

The echo Fe \Ka line flux can be related to the energy released by the putative illuminating source. In the optically thin regime the line photon flux $F_{\rm Fe}$ is \citep{Sunyaev+1998}:

\begin{equation}
\label{eq:reflection}
    F_{\rm Fe}= Y \frac{1}{\Delta t_{\rm cloud}}\frac{\Omega}{4 \pi D^2} \int \tau_{\rm Fe} (E) \, S(E,t) \, dE dt \quad \text{[photons s}^{-1} \text{cm}^{-2}\text{]},
\end{equation}
where $Y=0.3$ is the yield, $\Omega$ the solid angle under which the source sees the cloud, and $S(E,t)$ the photon spectrum. Since in this case the duration of the flare is comparable to or shorter than the light-crossing time of the clouds, we related the fluorescence signal to the total energy emitted by the source by integrating the luminosity over time. $\Delta t_{\rm cloud}\simeq 1.5$ yrs is the light-crossing time of the cloud. 
We assumed a $D=8.3$\,kpc distance from the GC \citep{Gravity_coolab_2019}. The optical depth, $\tau_{Fe}$, depends on the cross section of the iron K-shell, which we take as $\sigma_{Fe}=3.53 \times 10^{-20} \times (E/7.1 \, \text{keV})^{-3}$ cm$^2$ \citep{Sunyaev+1998}. For a given iron-to-hydrogen atom abundance ratio, $\delta_{Fe}$, the previous equation can be written as
\begin{equation}
    \label{eq:reflection_bis}
    \begin{split}
    F_{\rm Fe}=1.0\times10^{-6} \times \alpha \left(\frac{\Omega}{1.0}\right) \left(\frac{N_{\rm H_2}}{10^{23} \, \text{cm}^{-2}}\right) \left( \frac{\Delta t_{\rm cloud}}{1.5 \, \text{yrs}}\right)^{-1}  \\
    \times \left(\frac{\delta_{\rm Fe}}{5.9 \times 10^{-5}}\right)  \left(\frac{U_{2-10}\quad}{10^{42} \ \text{erg}} \right) 
 \quad \text{[photons s}^{-1} \text{cm}^{-2}\text{]},
 \end{split}
\end{equation}
where $U_{2-10}$ is the total energy released in the [2-10]\,keV band. We chose an iron abundance relative to hydrogen twice the solar one  (see \citet{Anastasopoulou+2023} and reference therein; the solar ratio is $2.95 \times 10^{-5}$ \citep{Lodders+2003}), assuming all the hydrogen is in a molecular state ($N_{\rm H_2}$ being the corresponding column density). Finally, $\alpha$ is the parameter accounting for the spectral shape of the source (see Appendix A for details), ranging from 0.3 to 3.4 keV $^{-1}$ for a black body with a temperature in the range [0.8-3]\,keV, and from 2.1 to 0.8 keV $^{-1}$ for a power law with a photon index in the range [1.5-3.0]. 

From $F_{\rm Fe}$ measured in 2019, 2021, and 2022, and from $U_{2-10}$ of the transient that generates the echo signal, eq.\ref{eq:reflection_bis} can be used to derive the combination $\Omega \times N_{\rm H_2}$. 
Then, in order to constrain the solid angle $\Omega$, we assumed for the CND geometry an inclination angle of 70$^\circ$. The thickness of the disk is estimated to be $\simeq 0.4$\,pc at the inner edge, located $\simeq 1.5$\,pc away from \sgra \citep{Jackson+1993}, increasing to $\simeq 2$\,pc at a distance of $\simeq 7$\,pc \citep{Vollmer+2001}. It corresponds to a $\simeq 15^\circ$ opening angle from the center (corresponding to \sgra's position). In Fig.\ref{fig:CND}, middle panel, we report the schematic model of the CND geometry that we assume in the following.

Depending on the position of the illuminating source, the reprocessed radiation reaches the observer at different times. 
This is depicted in Fig.\ref{fig:CND}, right panel. For a source at coordinates $(x,y,z)=(0,0,0)$ (the GC) with the $z$-axis coincident with the LOS, the points located on the paraboloid surface \citep{Sunyaev+1998} of equation 
\begin{equation}
    \frac{z}{c}=- \left[ \Delta t^2 - \left( \frac{x}{c}\right)^2- \left( \frac{y}{c}\right)^2\right]/(2\Delta t)
\end{equation}
experience the same time delay, $\Delta t$. In the sketch, two representative iso-delay paraboloids ($\Delta t=$ 0.5 and 9 years) are drawn. The time delay originates from the different
lengths the light travels to irradiate the cloud and then reach
the observer. In the same plot we show three light-paths corresponding to the source observer (solid line), the source western-part observer (dashed line), and source eastern-part observer (dotted-dashed line). This last path is about 8-9 light yrs longer than the dashed line path. In this scenario, the first light should appear in the western region about four to six months after the outburst. The opposite region should be illuminated about eight to nine years after the flaring event. 

The projected fluorescent wavefront is superluminal in the side pointing toward the observer because of the inclination angle of the CND (see Appendix B). For an angle of 70$^\circ$, ignoring the CND vertical thickness, the velocity of the projected fluorescent wavefront is $5.7c$ (1.7 pc yrs$^{-1}$) in the western part. On the other hand, it is just $0.18c$ (0.05 pc yrs$^{-1}$) in the eastern region. The projected wavefront scans the distribution of the
molecular clouds slowly in the eastern part and faster in the western segment. 
In order to illuminate the northern and southern parts of the CND, instead, light travels for $\sim 1.5$ pc in the orthogonal direction with respect to the LOS. Therefore, the signal is expected to reach us with a delay of about five years. 

The time delay between the outburst and the echo radiation constrains the position of the illuminated cloud with respect to that of the illuminating source, along the LOS. If the light-crossing time of the projected separation is shorter than the time delay then the illuminating source lies in front of the cloud. Vice versa, if the time delay is shorter than the light-crossing time of the projected separation then the source lies behind the cloud (see Appendix B for details).

\subsection{The magnetar as an illuminating source}
\begin{figure*}
    \centering
    \includegraphics[width=0.8\linewidth]{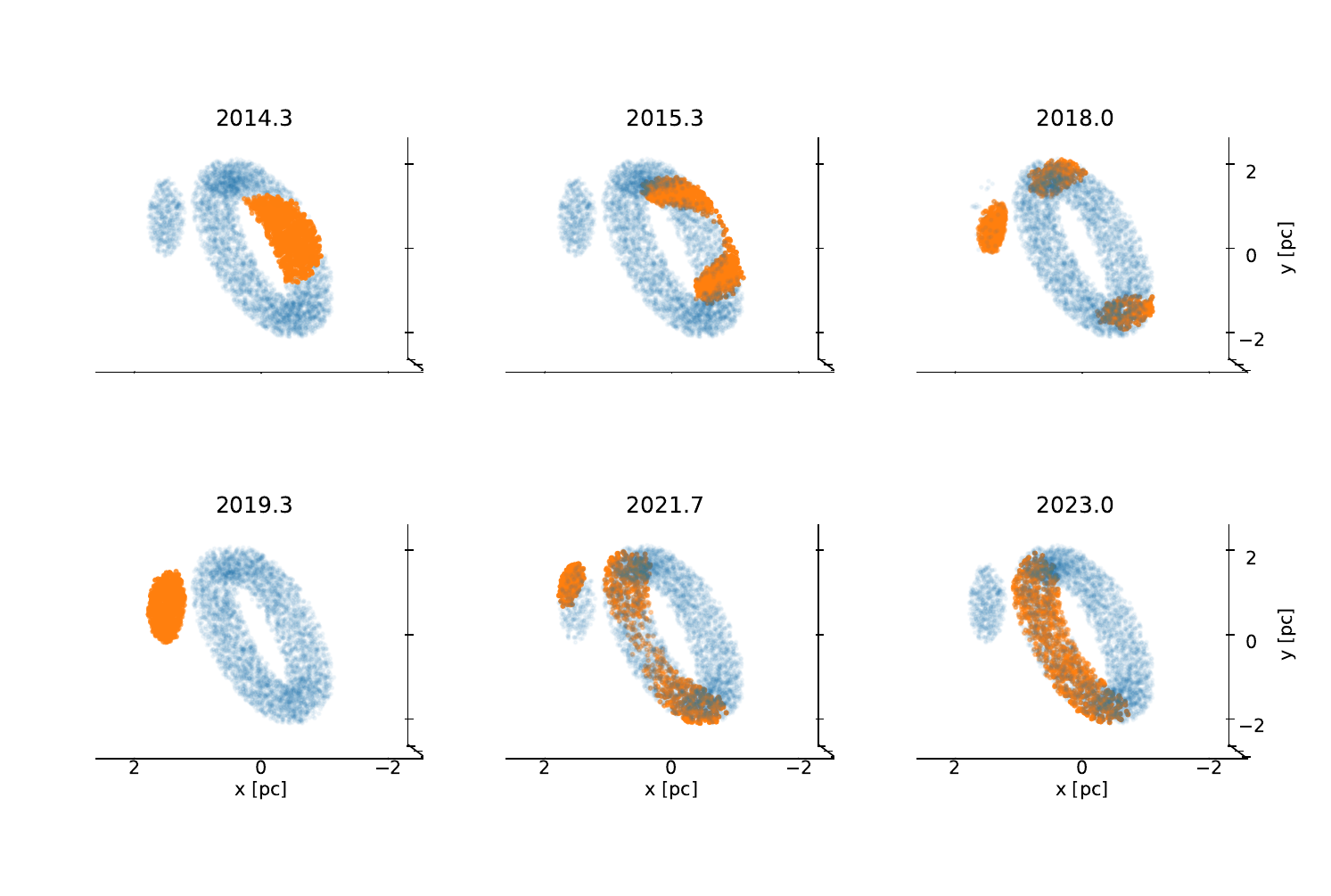}
    \caption{Qualitative evolution of the fluorescence signal from the CND (blue) due to the outburst of the magnetar SGR J1745-2900. The region from which a Fe Ka signal is expected is highlighted in orange. The density of the torus structure is assumed to be constant and uniform. We assume that the outburst had a one-year duration. The magnetar is assumed to be in the position $(x,y,z)=(0.1,-0.1,0.0)$ pc. A different position along the LOS translates in different time delays.} 
    \label{fig:cnd_model}
\end{figure*}

Fig.\ref{fig:cnd_model} displays a qualitative time evolution of the echo radiation from the CND in the case of the magnetar as the illuminating source. The region from which a Fe \Ka echo radiation is expected \footnote{An animated version is available at: \url{https://drive.google.com/file/d/1WTJnf5hjv0Mi5hZ9GaUm1cE4a5crFt25/view?usp=sharing} } is highlighted in orange. In this picture, we assume that the source is located 0.1 pc south and 0.1 east of \sgra, consistent with the measured position of SGR J1745-2900 \citep{Bower+2015}. Furthermore, we assume a uniform density of the torus structure and a duration of the outburst of one year. The signal may be more easily observed in the eastern region of the CND, where it lasts for a longer time and is distributed over a larger area. This double effect is due to the extended intrinsic geometry of the CND combined with the tilted angle along the LOS. In the northern and southern directions, instead, the amplitude of the wavefront of 1 light yr corresponds to an angular size of $\simeq 7"$ at the GC (less than two pixels in the images). The model provides a qualitative time evolution of the illuminating wavefront but does not take into account the clumpy nature of the CND. Moreover, the illuminating source is assumed to be in $z=0.0$ along the LOS. Small displacements from this position would result in a different time delay of the echo radiation. The magnetar position along the LOS is believed to be constrained by $\left|z\right|<0.1$\,pc \citep{Bower+2015}. It translates into uncertainties in the time delay of up to one year for different regions. Since region A also encloses part of the NEA, we included it in the model. The NEA is represented by the upper-left ellipsoid. Its position along the LOS was chosen so that a fluorescence signal was bound to be visible starting from 2019. Indeed, about $45"$ separates the NEA from \sgra, corresponding to a distance of $\simeq 1.8$\,pc at the GC position. If the emission observed in 2019 is due to the magnetar, then part of the NEA should be located at about the same location as the magnetar along the LOS, in front of the eastern part of the CND. In this case, the time delay is due to the travel distance orthogonal to the LOS of 1.8 pc ($\simeq 6$ light yrs). 

The signal found in the eastern section from 2019
is consistent with the scenario just described. On the other hand, the presence of the transient AX J1745.6-2901 together with the magnetar prevents us from studying the western region in the period 2013 to 2016, when a signal would be expected. 
Assuming a black-body ($T=0.8$\,keV) spectrum \citep{Rea+2020}, and using eq.\eqref{eq:reflection_bis} with the best estimate for $F_{\rm Fe}$, we derived $\Omega\times N_{\rm H_2} \simeq 1.5 \times 10^{23}$ sr cm$^{-2}$ for region A, where we adopted the total energy released by the magnetar during its first year. The light-crossing time of the clouds was assumed to be 1.5 yrs.
Assuming a 15$^\circ$ opening angle, and that the illuminated region is 1/6 of the entire CND circumference (for a total solid angle $\Omega=0.27$ sr), we derived for region A 
$N_{\rm H_2} =5 \times 10^{23}$ cm$^{-2}$, with a 68\% confidence interval of $[2.2; 9.6] \times 10^{23}$ cm$^{-2}$. The large uncertainties in the normalization of the Fe \Ka line are reflected in the large confidence interval. The found value is slightly higher but consistent with the values (a few $10^{22 \div 23} $ cm$^{-2}$) reported in the literature \citep{Latvakoski+1999,Etxaluze+2011, Mills+2017}.   

The confidence intervals account only for the uncertainty on the Fe \Ka line intensity. Other quantities can still modify the final column density estimate. As an example, the column density scales as $N_{\rm H_2} \propto \Delta t_{\rm cloud}\,\delta_{\rm Fe}^{-1} \, \Omega^{-1}$, so that the derived $N_{\rm H_2}$ decreases if one adopts a higher Fe abundance or a smaller light-crossing time of the clouds. 
Furthermore, the CND has a clumpy nature, with clouds $\lesssim 0.2$\,pc in scale. 
Also, the chosen solid angle takes into account only the proper disk. If the signal detected in 2019 is due to the magnetar, then some molecular material should be in front of the proper CND (such as the NEA in our model). Therefore, the radiation from the source is distributed over a larger $\Omega$. For example, for a solid angle, $\Omega=1.0$ sr, the derived interval for the column density shifts to $[0.6; 2.6] \times 10^{23}$ cm$^{-2}$.
Moreover, we related the fluorescence signal to the time-averaged flux of the illuminating source (in eq.\eqref{eq:reflection}, the spectrum is integrated over the emission time and divided by the mean light-crossing time of the cloud). The luminosity of the magnetar, in the 2-10 keV band, decreases from $4\times 10^{35}$ to $9 \times 10^{34}$ erg/s during the first year. In principle the estimate of $N_{\rm H_2}$ can be improved by future observations that directly relate the light curve of the illuminating source to the time variability of the line intensity.

\subsection{The Swift transients as illuminating sources}
As mentioned in Section \ref{sec:candidates}, other sources could contribute to the illuminating radiation, such as the two LMXBs that went into outburst during 2016. In this case the contribution of these transients must be added to that of the magnetar.

J174540.7-290015 is located only 32" from region A, which corresponds to a projected distance of 1.3 pc. If the binary is responsible for the fluorescence, then the echo delay is a distance
proxy from the illuminated cloud along the LOS. In this case, the time delay of 3.5 yrs is shorter than the light-crossing time of the projected distance (4.2 yrs). If the source illuminates the cloud in 2019 then it should be located behind the cloud (0.25 pc, see Appendix B for details). 
In this case $N_{\rm H_2}$ is constrained in $[5\times 10^{21}; 2\times 10^{22}]$ cm$^{-2}$ in region A, assuming a solid angle of $\Omega=1$ sr.

A similar argument can be applied to the other 2016 transient, that is to say, J174540.2-290037. In this case, the separation is $\simeq$ 40" (5.2 light yrs). Again, if the source were responsible for the fluorescent emission, it would be located behind the cloud illuminated in 2019 (0.65 pc). The higher distance would result in a smaller solid angle. For example, assuming $\Omega =0.3$\,sr, we derived a $N_{\rm H_2}$ in $[4\times 10^{22}; 2\times 10^{23}]$ cm$^{-2}$.

Even if these transients could, in principle, be responsible for
the fluorescence signal observed from 2019, it must be noted
that the primary candidate as an illuminating source remains the
magnetar SGR J1745-2900. Indeed, the magnetar generates a
fluorescence, given its position close to \sgra, and the signal
we observed is consistent both in time and space with what is
expected.

\section{Conclusions}
\label{sec:conclusions}
We studied the closest molecular clouds to \sgra\, located in the X-ray band of the CND. We investigated the variability of Fe \Ka line emission in the GC in order to associate the fluorescence emission with the outburst of one or more illuminating sources in the GC. We detect a variability of the Fe \Ka line intensity in a region consistent with the eastern part of the CND. The signal appears in 2019 and persists during 2021 and 2022. The typical intensity of the Fe \Ka line in this region for this period is a few $10^{-6}$ photons s$^{-1}$ cm$^{-2}$. We interpreted the Fe \Ka signal as an echo radiation from the magnetar SGR J1745-2900. From the geometry of the system, the Fe \Ka line intensity, and the measured energy released by the source, we infer the hydrogen column density ($N_{H_2}\sim 10^{23}$ cm $^{-2}$) of the CND on the eastern side, a value in agreement with the literature. There is strong evidence to support this scenario, since the measured fluorescence is observed where and when would be expected if the illuminating source were the magnetar, and the derived column density is in agreement with previous radio results.

Given the inclination angle ($70^\circ$) of the CND, the projected wavefront of the fluorescence has subluminal velocity in the eastern segment of the CND
and "slowly" scans the distribution of the molecular clouds. The signal should be visible in upcoming years. Multiple observations
over the years could directly compare the illuminating source light curve with the fluorescence light curve. In this way, one could study the response of the molecular clouds to the injected energy and better characterize the physical
properties of the clouds orbiting \sgra.

\begin{table*}[p]
\caption{Observation details: the set and observation number, the obs. ID, the starting time of the observations (UTC), and the exposure after filtering for high background flaring activity and \sgra's flares.}
\label{table:1}
\centering 
\begin{tabular}{cccccc}
\hline\hline
Set & obs. n. &obs. ID & Date & exposure & total exposure\\ 
    &         &       &       &  [s]     &     [s] \\        
\hline
\multirow{2}{*}{1} & 1 & 0202670501 & 2004-03-28T16:45:33 & 17400 &\multirow{2}{*}{50600} \\
 & 2 & 0202670601 & 2004-03-30T15:38:51 & 33200 & \\
\hline
\multirow{2}{*}{2} & 3 & 0202670701 & 2004-08-31T03:34:59 & 77600 &\multirow{2}{*}{167300} \\
& 4 & 0202670801 & 2004-09-02T03:24:36 & 89700 & \\
\hline
\multirow{3}{*}{3}
& 5 & 0554750401 & 2009-04-01T01:18:30 & 31600 & \multirow{3}{*}{100100}\\
& 6 & 0554750501 & 2009-04-03T01:56:31 & 37800 & \\
& 7 & 0554750601 & 2009-04-05T03:53:34 & 30700 & \\
\hline
\multirow{5}{*}{4} & 8 & 0604300601 & 2011-03-28T08:13:02 & 29100 & \multirow{5}{*}{147900}\\
& 9 & 0604300701 & 2011-03-30T09:25:58 & 32300 &\\
& 10 & 0604300801 & 2011-04-01T09:02:20 & 33900 &\\
& 11 & 0604300901 & 2011-04-03T08:15:10 & 19700 &\\
& 12 & 0604301001 & 2011-04-05T07:32:37 & 32900 &\\
\hline
\multirow{2}{*}{5} & 13 & 0658600101 & 2011-08-31T23:37:27 & 48000 &\multirow{2}{*}{88000} \\
& 14 & 0658600201 & 2011-09-01T20:26:54 & 40000 & \\
\hline
\multirow{5}{*}{6} &
15 & 0674600601 & 2012-03-13T04:15:33 & 8400 &\multirow{5}{*}{65900} \\
& 16 & 0674600701 & 2012-03-15T05:10:07 & 9400 &\\
& 17 & 0674600801 & 2012-03-19T04:15:37 & 16900 &\\
& 18 & 0674601001 & 2012-03-21T03:53:40 & 19800 &\\
& 19 & 0674601101 & 2012-03-17T02:53:14 & 11400 &\\
\hline
\multirow{3}{*}{7} &
20 & 0723410301 & 2014-02-28T18:21:51 & 18400 & \multirow{3}{*}{111100}\\
& 21 & 0723410401 & 2014-03-10T14:51:07 & 46200 &\\
& 22 & 0723410501 & 2014-04-02T03:41:13 & 46500 &\\
\hline
\multirow{4}{*}{8} & 23 & 0743630201 & 2014-08-30T20:00:24 & 21100 & \multirow{4}{*}{99300} \\
& 24 & 0743630301 & 2014-08-31T21:03:54 & 25100 &\\
& 25 & 0743630401 & 2014-09-27T19:47:57 & 15500 &\\
& 26 & 0743630501 & 2014-09-28T21:42:09 & 37600 &\\
\hline
\multirow{10}{*}{9} & 27 & 0822680201 & 2019-03-29T05:24:29 & 25500 & \multirow{10}{*}{163000}\\
& 28 & 0822680401 & 2019-03-31T05:27:07 & 30600 &\\
& 29 & 0822680501 & 2019-04-01T05:19:52 & 21000 &\\
& 30 & 0831800201 & 2019-03-12T00:44:36 & 11700 &\\
& 31 & 0831800301 & 2019-03-13T01:21:53 & 10800 &\\
& 32 & 0831800401 & 2019-03-14T00:54:24 & 10700 &\\
& 33 & 0831800501 & 2019-03-15T01:13:16 & 11600 &\\
& 34 & 0831800701 & 2019-03-17T01:05:46 & 11500 &\\
& 35 & 0831800801 & 2019-03-18T01:10:22 & 10800 &\\
& 36 & 0831801001 & 2019-04-03T23:53:21 & 18800 &\\
\hline
\multirow{2}{*}{10} & 37 & 0831801301 & 2019-09-01T18:44:01 & 11000 & \multirow{2}{*}{64500}\\
& 38 & 0851180901 & 2019-08-31T22:51:22 & 53500 & \\
\hline
\multirow{4}{*}{11} & 39 & 0860620201 & 2021-03-28T07:36:33 & 24500 & \multirow{4}{*}{70700}\\
& 40 & 0860620301 & 2021-03-29T06:40:23 & 8000 & \\
& 41 & 0860620401 & 2021-03-30T05:42:23 & 27400 & \\
& 42 & 0860620501 & 2021-03-31T05:42:24 & 10800 & \\
\hline
\multirow{2}{*}{12} & 43 & 0893811101 & 2022-03-18T01:15:45 & 49900 & \multirow{2}{*}{102700}\\
& 44 & 0893811301 & 2022-03-26T00:48:15 & 52800 & \\
\hline
\end{tabular}
\end{table*}

\begin{acknowledgements}
      We thank Konrad Dennerl and the XMM-Newton Helpdesk for the helpful discussions. This work is based on observations obtained with XMM-Newton, an ESA science mission with instruments and contributions directly funded by ESA Member States and NASA. This project acknowledges funding from the European Research Council (ERC) under the European Union’s Horizon
2020 research and innovation programme (grant agreement No
865637), and support from Bando per il Finanziamento della Ricerca Fondamentale 2022 dell’Istituto Nazionale di Astrofisica (INAF): GO Large program.
\end{acknowledgements}

\bibliography{sample}

\begin{thebibliography}{69}
\expandafter\ifx\csname natexlab\endcsname\relax\def\natexlab#1{#1}\fi

\bibitem[{{Anastasopoulou} {et~al.}(2023){Anastasopoulou}, {Ponti}, {Sormani},
  {Locatelli}, {Haberl}, {Morris}, {Churazov}, {Sch{\"o}del}, {Maitra},
  {Campana}, {Di Teodoro}, {Jin}, {Khabibullin}, {Mondal}, {Sasaki}, {Zhang},
  \& {Zheng}}]{Anastasopoulou+2023}
{Anastasopoulou}, K., {Ponti}, G., {Sormani}, M.~C., {et~al.} 2023, \aap, 671,
  A55

\bibitem[{{Arnaud}(1996)}]{Arnaud1996}
{Arnaud}, K.~A. 1996, in Astronomical Society of the Pacific Conference Series,
  Vol. 101, Astronomical Data Analysis Software and Systems V, ed. G.~H.
  {Jacoby} \& J.~{Barnes}, 17

\bibitem[{{Baganoff} {et~al.}(2001){Baganoff}, {Bautz}, {Brandt}, {Chartas},
  {Feigelson}, {Garmire}, {Maeda}, {Morris}, {Ricker}, {Townsley}, \&
  {Walter}}]{Baganoff+2001}
{Baganoff}, F.~K., {Bautz}, M.~W., {Brandt}, W.~N., {et~al.} 2001, \nat, 413,
  45

\bibitem[{{Baganoff} {et~al.}(2003){Baganoff}, {Maeda}, {Morris}, {Bautz},
  {Brandt}, {Cui}, {Doty}, {Feigelson}, {Garmire}, {Pravdo}, {Ricker}, \&
  {Townsley}}]{Baganoff+2003}
{Baganoff}, F.~K., {Maeda}, Y., {Morris}, M., {et~al.} 2003, \apj, 591, 891

\bibitem[{Bocquet \& Carter(2016)}]{Bocquet2016}
Bocquet, S. \& Carter, F.~W. 2016, The Journal of Open Source Software, 1

\bibitem[{{Bower} {et~al.}(2015){Bower}, {Deller}, {Demorest}, {Brunthaler},
  {Falcke}, {Moscibrodzka}, {O'Leary}, {Eatough}, {Kramer}, {Lee}, {Spitler},
  {Desvignes}, {Rushton}, {Doeleman}, \& {Reid}}]{Bower+2015}
{Bower}, G.~C., {Deller}, A., {Demorest}, P., {et~al.} 2015, \apj, 798, 120

\bibitem[{{Buchner} {et~al.}(2014){Buchner}, {Georgakakis}, {Nandra}, {Hsu},
  {Rangel}, {Brightman}, {Merloni}, {Salvato}, {Donley}, \&
  {Kocevski}}]{Buchner+2014}
{Buchner}, J., {Georgakakis}, A., {Nandra}, K., {et~al.} 2014, \aap, 564, A125

\bibitem[{{Christopher} {et~al.}(2005){Christopher}, {Scoville}, {Stolovy}, \&
  {Yun}}]{Cristopher+2005}
{Christopher}, M.~H., {Scoville}, N.~Z., {Stolovy}, S.~R., \& {Yun}, M.~S.
  2005, \apj, 622, 346

\bibitem[{{Chuard} {et~al.}(2018){Chuard}, {Terrier}, {Goldwurm}, {Clavel},
  {Soldi}, {Morris}, {Ponti}, {Walls}, \& {Chernyakova}}]{Chuard+2018}
{Chuard}, D., {Terrier}, R., {Goldwurm}, A., {et~al.} 2018, \aap, 610, A34

\bibitem[{{Churazov} {et~al.}(2017){Churazov}, {Khabibullin}, {Sunyaev}, \&
  {Ponti}}]{Chrazov+2017}
{Churazov}, E., {Khabibullin}, I., {Sunyaev}, R., \& {Ponti}, G. 2017, \mnras,
  465, 45

\bibitem[{{Clavel} {et~al.}(2013){Clavel}, {Terrier}, {Goldwurm}, {Morris},
  {Ponti}, {Soldi}, \& {Trap}}]{Clavel+2013}
{Clavel}, M., {Terrier}, R., {Goldwurm}, A., {et~al.} 2013, \aap, 558, A32

\bibitem[{{Corrales} {et~al.}(2017){Corrales}, {Mon}, {Haggard}, {Baganoff},
  {Garmire}, {Degenaar}, \& {Reynolds}}]{Corrales+2017}
{Corrales}, L.~R., {Mon}, B., {Haggard}, D., {et~al.} 2017, \apj, 839, 76

\bibitem[{{Coti Zelati} {et~al.}(2015){Coti Zelati}, {Rea}, {Papitto},
  {Vigan{\`o}}, {Pons}, {Turolla}, {Esposito}, {Haggard}, {Baganoff}, {Ponti},
  {Israel}, {Campana}, {Torres}, {Tiengo}, {Mereghetti}, {Perna}, {Zane},
  {Mignani}, {Possenti}, \& {Stella}}]{CotiZelati+2015}
{Coti Zelati}, F., {Rea}, N., {Papitto}, A., {et~al.} 2015, \mnras, 449, 2685

\bibitem[{{Coti Zelati} {et~al.}(2017){Coti Zelati}, {Rea}, {Turolla}, {Pons},
  {Papitto}, {Esposito}, {Israel}, {Campana}, {Zane}, {Tiengo}, {Mignani},
  {Mereghetti}, {Baganoff}, {Haggard}, {Ponti}, {Torres}, {Borghese}, \&
  {Elfritz}}]{CotiZelati+2017}
{Coti Zelati}, F., {Rea}, N., {Turolla}, R., {et~al.} 2017, \mnras, 471, 1819

\bibitem[{{Couderc}(1939)}]{Couderc1939}
{Couderc}, P. 1939, Annales d'Astrophysique, 2, 271

\bibitem[{{Degenaar} {et~al.}(2013){Degenaar}, {Reynolds}, {Miller}, {Kennea},
  \& {Wijnands}}]{Degenaar+2013}
{Degenaar}, N., {Reynolds}, M.~T., {Miller}, J.~M., {Kennea}, J.~A., \&
  {Wijnands}, R. 2013, The Astronomer's Telegram, 5006, 1

\bibitem[{{Degenaar} {et~al.}(2016{\natexlab{a}}){Degenaar}, {Reynolds},
  {Wijnands}, {Miller}, {Kennea}, {Ponti}, {Haggard}, \&
  {Gehrels}}]{Degenaar+2016b}
{Degenaar}, N., {Reynolds}, M.~T., {Wijnands}, R., {et~al.} 2016{\natexlab{a}},
  The Astronomer's Telegram, 9196, 1

\bibitem[{{Degenaar} {et~al.}(2016{\natexlab{b}}){Degenaar}, {Reynolds},
  {Wijnands}, {Miller}, {Kennea}, {Ponti}, {Haggard}, \&
  {Gehrels}}]{Degenaar+2016}
{Degenaar}, N., {Reynolds}, M.~T., {Wijnands}, R., {et~al.} 2016{\natexlab{b}},
  The Astronomer's Telegram, 9109, 1

\bibitem[{{Degenaar} {et~al.}(2015){Degenaar}, {Wijnands}, {Miller},
  {Reynolds}, {Kennea}, \& {Gehrels}}]{Degenaar+2015}
{Degenaar}, N., {Wijnands}, R., {Miller}, J.~M., {et~al.} 2015, Journal of High
  Energy Astrophysics, 7, 137

\bibitem[{{Dinh} {et~al.}(2021){Dinh}, {Salas}, {Morris}, \&
  {Naoz}}]{Dinh+2021}
{Dinh}, C.~K., {Salas}, J.~M., {Morris}, M.~R., \& {Naoz}, S. 2021, \apj, 920,
  79

\bibitem[{{Etxaluze} {et~al.}(2011){Etxaluze}, {Smith}, {Tolls}, {Stark}, \&
  {Gonz{\'a}lez-Alfonso}}]{Etxaluze+2011}
{Etxaluze}, M., {Smith}, H.~A., {Tolls}, V., {Stark}, A.~A., \&
  {Gonz{\'a}lez-Alfonso}, E. 2011, \aj, 142, 134

\bibitem[{{Genzel}(1989)}]{Genzel1989}
{Genzel}, R. 1989, in The Center of the Galaxy, ed. M.~{Morris}, Vol. 136, 393

\bibitem[{{Genzel} {et~al.}(2010){Genzel}, {Eisenhauer}, \&
  {Gillessen}}]{Genzel+2010}
{Genzel}, R., {Eisenhauer}, F., \& {Gillessen}, S. 2010, Reviews of Modern
  Physics, 82, 3121

\bibitem[{{GRAVITY Collaboration} {et~al.}(2019){GRAVITY Collaboration},
  {Abuter}, {Amorim}, {Baub{\"o}ck}, {Berger}, {Bonnet}, {Brandner},
  {Cl{\'e}net}, {Coud{\'e} Du Foresto}, {de Zeeuw}, {Dexter}, {Duvert},
  {Eckart}, {Eisenhauer}, {F{\"o}rster Schreiber}, {Garcia}, {Gao}, {Gendron},
  {Genzel}, {Gerhard}, {Gillessen}, {Habibi}, {Haubois}, {Henning}, {Hippler},
  {Horrobin}, {Jim{\'e}nez-Rosales}, {Jocou}, {Kervella}, {Lacour},
  {Lapeyr{\`e}re}, {Le Bouquin}, {L{\'e}na}, {Ott}, {Paumard}, {Perraut},
  {Perrin}, {Pfuhl}, {Rabien}, {Rodriguez Coira}, {Rousset}, {Scheithauer},
  {Sternberg}, {Straub}, {Straubmeier}, {Sturm}, {Tacconi}, {Vincent}, {von
  Fellenberg}, {Waisberg}, {Widmann}, {Wieprecht}, {Wiezorrek}, {Woillez}, \&
  {Yazici}}]{Gravity_coolab_2019}
{GRAVITY Collaboration}, {Abuter}, R., {Amorim}, A., {et~al.} 2019, \aap, 625,
  L10

\bibitem[{{Haggard} {et~al.}(2019){Haggard}, {Nynka}, {Mon}, {de la Cruz
  Hernandez}, {Nowak}, {Heinke}, {Neilsen}, {Dexter}, {Fragile}, {Baganoff},
  {Bower}, {Corrales}, {Coti Zelati}, {Degenaar}, {Markoff}, {Morris}, {Ponti},
  {Rea}, {Wilms}, \& {Yusef-Zadeh}}]{Haggard+2019}
{Haggard}, D., {Nynka}, M., {Mon}, B., {et~al.} 2019, \apj, 886, 96

\bibitem[{{Hsieh} {et~al.}(2017){Hsieh}, {Koch}, {Ho}, {Kim}, {Tang}, {Wang},
  {Yen}, \& {Hwang}}]{Hsieh+2017}
{Hsieh}, P.-Y., {Koch}, P.~M., {Ho}, P. T.~P., {et~al.} 2017, \apj, 847, 3

\bibitem[{{Hsieh} {et~al.}(2021){Hsieh}, {Koch}, {Kim}, {Mart{\'\i}n}, {Yen},
  {Carpenter}, {Harada}, {Turner}, {Ho}, {Tang}, \& {Beck}}]{Hsieh+2021}
{Hsieh}, P.-Y., {Koch}, P.~M., {Kim}, W.-T., {et~al.} 2021, \apj, 913, 94

\bibitem[{{Inui} {et~al.}(2009){Inui}, {Koyama}, {Matsumoto}, \&
  {Tsuru}}]{Inui+2009}
{Inui}, T., {Koyama}, K., {Matsumoto}, H., \& {Tsuru}, T.~G. 2009, \pasj, 61,
  S241

\bibitem[{{Jackson} {et~al.}(1993){Jackson}, {Geis}, {Genzel}, {Harris},
  {Madden}, {Poglitsch}, {Stacey}, \& {Townes}}]{Jackson+1993}
{Jackson}, J.~M., {Geis}, N., {Genzel}, R., {et~al.} 1993, \apj, 402, 173

\bibitem[{{Jeffreys}(1946)}]{Jeffreys+1946}
{Jeffreys}, H. 1946, Proceedings of the Royal Society of London Series A, 186,
  453

\bibitem[{{Kapteyn}(1902)}]{Kapteyn1902}
{Kapteyn}, J.~C. 1902, Popular Astronomy, 10, 124

\bibitem[{{Kennea} {et~al.}(2013){Kennea}, {Burrows}, {Kouveliotou}, {Palmer},
  {G{\"o}{\u{g}}{\"u}{\c{s}}}, {Kaneko}, {Evans}, {Degenaar}, {Reynolds},
  {Miller}, {Wijnands}, {Mori}, \& {Gehrels}}]{Kennea+2013}
{Kennea}, J.~A., {Burrows}, D.~N., {Kouveliotou}, C., {et~al.} 2013, \apjl,
  770, L24

\bibitem[{{Koyama} {et~al.}(1996){Koyama}, {Maeda}, {Sonobe}, {Takeshima},
  {Tanaka}, \& {Yamauchi}}]{Koyama+1996}
{Koyama}, K., {Maeda}, Y., {Sonobe}, T., {et~al.} 1996, \pasj, 48, 249

\bibitem[{{Latvakoski} {et~al.}(1999){Latvakoski}, {Stacey}, {Gull}, \&
  {Hayward}}]{Latvakoski+1999}
{Latvakoski}, H.~M., {Stacey}, G.~J., {Gull}, G.~E., \& {Hayward}, T.~L. 1999,
  \apj, 511, 761

\bibitem[{{Lau} {et~al.}(2013){Lau}, {Herter}, {Morris}, {Becklin}, \&
  {Adams}}]{Lau+2013}
{Lau}, R.~M., {Herter}, T.~L., {Morris}, M.~R., {Becklin}, E.~E., \& {Adams},
  J.~D. 2013, \apj, 775, 37

\bibitem[{{Lodders}(2003)}]{Lodders+2003}
{Lodders}, K. 2003, \apj, 591, 1220

\bibitem[{{Maeda} {et~al.}(2002){Maeda}, {Baganoff}, {Feigelson}, {Morris},
  {Bautz}, {Brandt}, {Burrows}, {Doty}, {Garmire}, {Pravdo}, {Ricker}, \&
  {Townsley}}]{Maeda+2002}
{Maeda}, Y., {Baganoff}, F.~K., {Feigelson}, E.~D., {et~al.} 2002, \apj, 570,
  671

\bibitem[{Marin {et~al.}(2023)Marin, Churazov, Khabibullin, Ferrazzoli,
  Di~Gesu, Barnouin, Di~Marco, Middei, Vikhlinin, Costa, Soffitta, Muleri,
  Sunyaev, Forman, Kraft, Bianchi, Donnarumma, Petrucci, Enoto, Agudo,
  Antonelli, Bachetti, Baldini, Baumgartner, Bellazzini, Bongiorno, Bonino,
  Brez, Bucciantini, Capitanio, Castellano, Cavazzuti, Chen, Ciprini, De~Rosa,
  Del~Monte, Di~Lalla, Doroshenko, Dov{\v c}iak, Ehlert, Evangelista, Fabiani,
  Garcia, Gunji, Hayashida, Heyl, Ingram, Iwakiri, Jorstad, Kaaret, Karas,
  Kitaguchi, Kolodziejczak, Krawczynski, La~Monaca, Latronico, Liodakis,
  Maldera, Manfreda, Marinucci, Marscher, Marshall, Massaro, Matt, Mitsuishi,
  Mizuno, Negro, Ng, O'Dell, Omodei, Oppedisano, Papitto, Pavlov, Peirson,
  Perri, Pesce-Rollins, Pilia, Possenti, Poutanen, Puccetti, Ramsey, Rankin,
  Ratheesh, Roberts, Romani, Sgr{\`o}, Slane, Spandre, Swartz, Tamagawa,
  Tavecchio, Taverna, Tawara, Tennant, Thomas, Tombesi, Trois, Tsygankov,
  Turolla, Vink, Weisskopf, Wu, Xie, \& Zane}]{Marin+2023}
Marin, F., Churazov, E., Khabibullin, I., {et~al.} 2023, Nature, 619, 41

\bibitem[{{Markevitch} {et~al.}(1993){Markevitch}, {Sunyaev}, \&
  {Pavlinsky}}]{Markevitch+1993}
{Markevitch}, M., {Sunyaev}, R.~A., \& {Pavlinsky}, M. 1993, \nat, 364, 40

\bibitem[{{Mart{\'\i}n} {et~al.}(2012){Mart{\'\i}n}, {Mart{\'\i}n-Pintado},
  {Montero-Casta{\~n}o}, {Ho}, \& {Blundell}}]{Martin+2012}
{Mart{\'\i}n}, S., {Mart{\'\i}n-Pintado}, J., {Montero-Casta{\~n}o}, M., {Ho},
  P.~T.~P., \& {Blundell}, R. 2012, \aap, 539, A29

\bibitem[{{Mills} {et~al.}(2017){Mills}, {Togi}, \& {Kaufman}}]{Mills+2017}
{Mills}, E. A.~C., {Togi}, A., \& {Kaufman}, M. 2017, \apj, 850, 192

\bibitem[{{Mori} {et~al.}(2013){Mori}, {Gotthelf}, {Zhang}, {An}, {Baganoff},
  {Barri{\`e}re}, {Beloborodov}, {Boggs}, {Christensen}, {Craig}, {Dufour},
  {Grefenstette}, {Hailey}, {Harrison}, {Hong}, {Kaspi}, {Kennea}, {Madsen},
  {Markwardt}, {Nynka}, {Stern}, {Tomsick}, \& {Zhang}}]{Mori+2013}
{Mori}, K., {Gotthelf}, E.~V., {Zhang}, S., {et~al.} 2013, \apjl, 770, L23

\bibitem[{{Mori} {et~al.}(2019){Mori}, {Hailey}, {Mandel}, {Schutt},
  {Bachetti}, {Coerver}, {Baganoff}, {Dykaar}, {Grindlay}, {Haggard}, {Heuer},
  {Hong}, {Hord}, {Jin}, {Nynka}, {Ponti}, \& {Tomsick}}]{Mori+2019}
{Mori}, K., {Hailey}, C.~J., {Mandel}, S., {et~al.} 2019, \apj, 885, 142

\bibitem[{{Morris} \& {Serabyn}(1996)}]{Morris+1996}
{Morris}, M. \& {Serabyn}, E. 1996, \araa, 34, 645

\bibitem[{{Mossoux} \& {Eckart}(2018)}]{Mossoux+2018}
{Mossoux}, E. \& {Eckart}, A. 2018, \mnras, 474, 3787

\bibitem[{{Muno} {et~al.}(2003){Muno}, {Baganoff}, {Bautz}, {Brandt}, {Broos},
  {Feigelson}, {Garmire}, {Morris}, {Ricker}, \& {Townsley}}]{Muno+2003}
{Muno}, M.~P., {Baganoff}, F.~K., {Bautz}, M.~W., {et~al.} 2003, \apj, 589, 225

\bibitem[{{Muno} {et~al.}(2007){Muno}, {Baganoff}, {Brandt}, {Park}, \&
  {Morris}}]{Muno+2007}
{Muno}, M.~P., {Baganoff}, F.~K., {Brandt}, W.~N., {Park}, S., \& {Morris},
  M.~R. 2007, \apjl, 656, L69

\bibitem[{{Muno} {et~al.}(2005){Muno}, {Lu}, {Baganoff}, {Brandt}, {Garmire},
  {Ghez}, {Hornstein}, \& {Morris}}]{Muno+2005}
{Muno}, M.~P., {Lu}, J.~R., {Baganoff}, F.~K., {et~al.} 2005, \apj, 633, 228

\bibitem[{{Neilsen} {et~al.}(2013){Neilsen}, {Nowak}, {Gammie}, {Dexter},
  {Markoff}, {Haggard}, {Nayakshin}, {Wang}, {Grosso}, {Porquet}, {Tomsick},
  {Degenaar}, {Fragile}, {Houck}, {Wijnands}, {Miller}, \&
  {Baganoff}}]{Neilsen+2013}
{Neilsen}, J., {Nowak}, M.~A., {Gammie}, C., {et~al.} 2013, \apj, 774, 42

\bibitem[{{Nobukawa} {et~al.}(2011){Nobukawa}, {Ryu}, {Tsuru}, \&
  {Koyama}}]{Nobukawa+2011}
{Nobukawa}, M., {Ryu}, S.~G., {Tsuru}, T.~G., \& {Koyama}, K. 2011, \apjl, 739,
  L52

\bibitem[{{Ponti} {et~al.}(2015){Ponti}, {De Marco}, {Morris}, {Merloni},
  {Mu{\~n}oz-Darias}, {Clavel}, {Haggard}, {Zhang}, {Nandra}, {Gillessen},
  {Mori}, {Neilsen}, {Rea}, {Degenaar}, {Terrier}, \& {Goldwurm}}]{Ponti+2015}
{Ponti}, G., {De Marco}, B., {Morris}, M.~R., {et~al.} 2015, \mnras, 454, 1525

\bibitem[{{Ponti} {et~al.}(2016){Ponti}, {Jin}, {De Marco}, {Rea}, {Rau},
  {Haberl}, {Coti Zelati}, {Bozzo}, {Ferrigno}, {Bower}, \&
  {Demorest}}]{Ponti+2016}
{Ponti}, G., {Jin}, C., {De Marco}, B., {et~al.} 2016, \mnras, 461, 2688

\bibitem[{{Ponti} {et~al.}(2013){Ponti}, {Morris}, {Terrier}, \&
  {Goldwurm}}]{Ponti+2013}
{Ponti}, G., {Morris}, M.~R., {Terrier}, R., \& {Goldwurm}, A. 2013, in
  Astrophysics and Space Science Proceedings, Vol.~34, Cosmic Rays in
  Star-Forming Environments, ed. D.~F. {Torres} \& O.~{Reimer}, 331

\bibitem[{{Ponti} {et~al.}(2010){Ponti}, {Terrier}, {Goldwurm}, {Belanger}, \&
  {Trap}}]{Ponti+2010}
{Ponti}, G., {Terrier}, R., {Goldwurm}, A., {Belanger}, G., \& {Trap}, G. 2010,
  \apj, 714, 732

\bibitem[{{Porquet} {et~al.}(2005){Porquet}, {Grosso}, {B{\'e}langer},
  {Goldwurm}, {Yusef-Zadeh}, {Warwick}, \& {Predehl}}]{Porquet+2005}
{Porquet}, D., {Grosso}, N., {B{\'e}langer}, G., {et~al.} 2005, \aap, 443, 571

\bibitem[{{Porquet} {et~al.}(2003){Porquet}, {Predehl}, {Aschenbach}, {Grosso},
  {Goldwurm}, {Goldoni}, {Warwick}, \& {Decourchelle}}]{Porquet+2003}
{Porquet}, D., {Predehl}, P., {Aschenbach}, B., {et~al.} 2003, \aap, 407, L17

\bibitem[{{Rea} {et~al.}(2020){Rea}, {Coti Zelati}, {Vigan{\`o}}, {Papitto},
  {Baganoff}, {Borghese}, {Campana}, {Esposito}, {Haggard}, {Israel},
  {Mereghetti}, {Mignani}, {Perna}, {Pons}, {Ponti}, {Stella}, {Torres},
  {Turolla}, \& {Zane}}]{Rea+2020}
{Rea}, N., {Coti Zelati}, F., {Vigan{\`o}}, D., {et~al.} 2020, \apj, 894, 159

\bibitem[{{Rea} {et~al.}(2013){Rea}, {Esposito}, {Pons}, {Turolla}, {Torres},
  {Israel}, {Possenti}, {Burgay}, {Vigan{\`o}}, {Papitto}, {Perna}, {Stella},
  {Ponti}, {Baganoff}, {Haggard}, {Camero-Arranz}, {Zane}, {Minter},
  {Mereghetti}, {Tiengo}, {Sch{\"o}del}, {Feroci}, {Mignani}, \&
  {G{\"o}tz}}]{Rea+2013}
{Rea}, N., {Esposito}, P., {Pons}, J.~A., {et~al.} 2013, \apjl, 775, L34

\bibitem[{{Requena-Torres} {et~al.}(2012){Requena-Torres}, {G{\"u}sten},
  {Wei{\ss}}, {Harris}, {Mart{\'\i}n-Pintado}, {Stutzki}, {Klein}, {Heyminck},
  \& {Risacher}}]{Torres+2012}
{Requena-Torres}, M.~A., {G{\"u}sten}, R., {Wei{\ss}}, A., {et~al.} 2012, \aap,
  542, L21

\bibitem[{{Reynolds} {et~al.}(2016){Reynolds}, {Kennea}, {Degenaar},
  {Wijnands}, \& {Miller}}]{Reynolds+2016}
{Reynolds}, M., {Kennea}, J., {Degenaar}, N., {Wijnands}, R., \& {Miller}, J.
  2016, The Astronomer's Telegram, 8649, 1

\bibitem[{SOC(2022)}]{XMM_manual}
SOC, X.-N. 2022, Users Guide to the XMM-Newton Science Analysis System, issue
  17.0

\bibitem[{{Sunyaev} \& {Churazov}(1998)}]{Sunyaev+1998}
{Sunyaev}, R. \& {Churazov}, E. 1998, \mnras, 297, 1279

\bibitem[{{Sunyaev} {et~al.}(1993){Sunyaev}, {Markevitch}, \&
  {Pavlinsky}}]{Sunyaev+1993}
{Sunyaev}, R.~A., {Markevitch}, M., \& {Pavlinsky}, M. 1993, \apj, 407, 606

\bibitem[{{Terrier} {et~al.}(2018){Terrier}, {Clavel}, {Soldi}, {Goldwurm},
  {Ponti}, {Morris}, \& {Chuard}}]{Terrier+2018}
{Terrier}, R., {Clavel}, M., {Soldi}, S., {et~al.} 2018, \aap, 612, A102

\bibitem[{{Tsuboi} {et~al.}(2018){Tsuboi}, {Kitamura}, {Uehara}, {Tsutsumi},
  {Miyawaki}, {Miyoshi}, \& {Miyazaki}}]{Tsuboi+2018}
{Tsuboi}, M., {Kitamura}, Y., {Uehara}, K., {et~al.} 2018, \pasj, 70, 85

\bibitem[{{Vollmer} \& {Duschl}(2001)}]{Vollmer+2001}
{Vollmer}, B. \& {Duschl}, W.~J. 2001, \aap, 367, 72

\bibitem[{{Wright} {et~al.}(2001){Wright}, {Coil}, {McGary}, {Ho}, \&
  {Harris}}]{Wright+2001}
{Wright}, M. C.~H., {Coil}, A.~L., {McGary}, R.~S., {Ho}, P. T.~P., \&
  {Harris}, A.~I. 2001, \apj, 551, 254

\bibitem[{{Zhang} {et~al.}(2015){Zhang}, {Hailey}, {Mori}, {Clavel}, {Terrier},
  {Ponti}, {Goldwurm}, {Bauer}, {Boggs}, {Christensen}, {Craig}, {Harrison},
  {Hong}, {Nynka}, {Soldi}, {Stern}, {Tomsick}, \& {Zhang}}]{Zhang+2015}
{Zhang}, S., {Hailey}, C.~J., {Mori}, K., {et~al.} 2015, \apj, 815, 132

\bibitem[{{Zhao} {et~al.}(2016){Zhao}, {Morris}, \& {Goss}}]{Zhao+2016}
{Zhao}, J.-H., {Morris}, M.~R., \& {Goss}, W.~M. 2016, \apj, 817, 171

\end{thebibliography}

\begin{appendix}
\section{}
\begin{figure}[]
    \centering
    \includegraphics[width=0.9\linewidth]{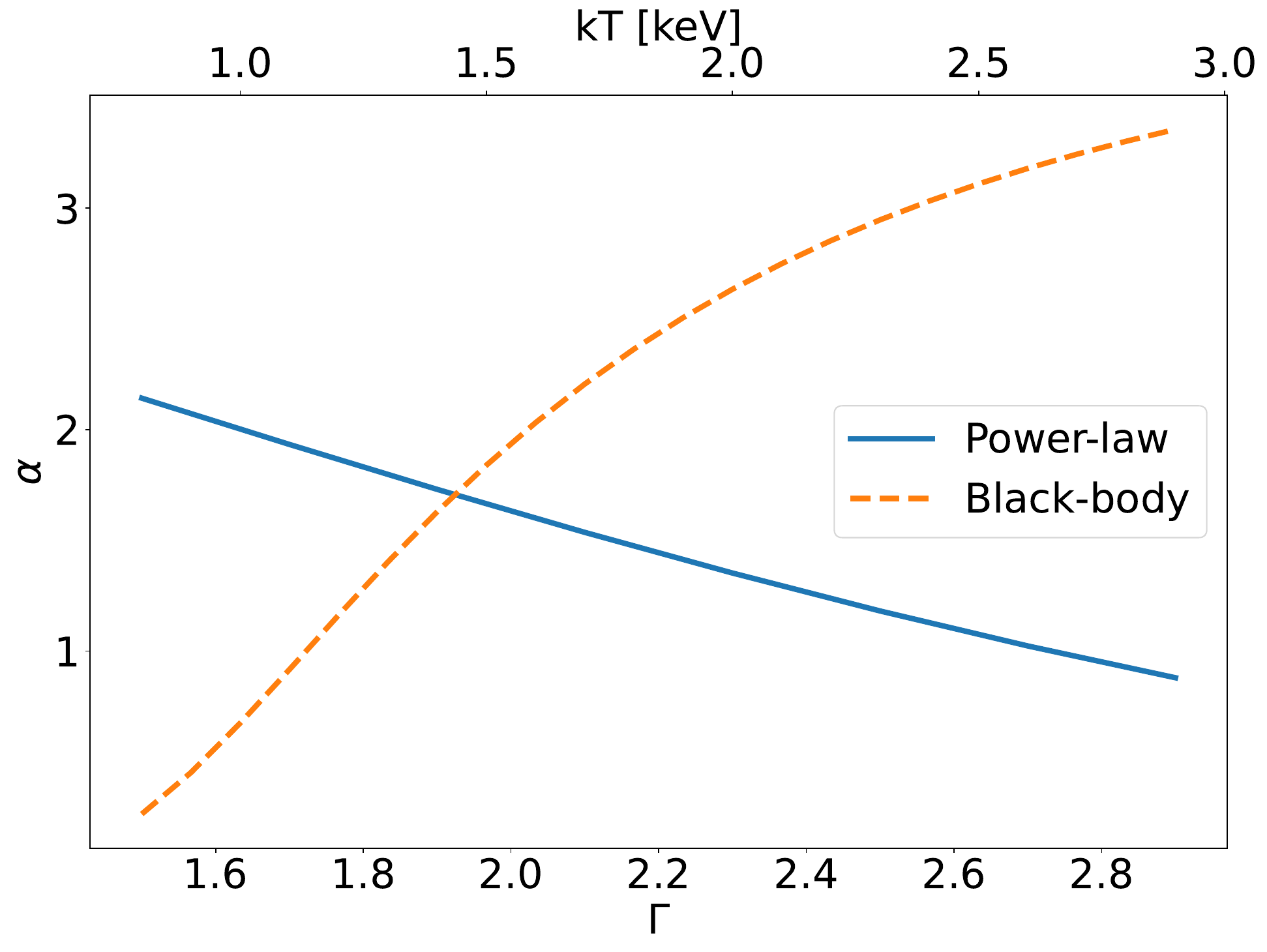}
    \caption{$\alpha$ parameter as a function of the temperature (black-body, dashed line) and the photon index (power-law, solid line).}
    \label{fig:alpha}
\end{figure}

Starting from eq.\eqref{eq:reflection}, and including the spectral dependence of the Fe cross section, the 6.4 kev line intensity is
\begin{equation}
\begin{split}
    F_{Fe}= Y \frac{1}{\Delta t_{cloud}}\frac{\Omega}{4 \pi D^2} N_{Fe} \sigma_{Fe} (7.1) \times \int_{7.1}^{\infty}  \, S(E,t) \, (E/7.1)^{-3}\, dE \, dt,
\end{split}
\end{equation}
where the spectrum is integrated over time, since in this case the flare duration is comparable to or shorter than the light-crossing time of the cloud. We defined $\alpha$, the ratio, as
\begin{equation}
\begin{split}
    \alpha \equiv 10^{2} \times \frac{\int_{7.1}^{\infty}  S(E,t) \,(E/7.1)^{-3} \, dE }{ \int_2^{10} E \, S(E,t) \, dE} \qquad [\text{keV}^{-1}],
\end{split}
\end{equation}
where the denominator is the luminosity in the 2-10 keV band.
The parameter, $\alpha$, depends on the spectral shape of the source. Fig. \ref{fig:alpha} displays $\alpha$ for common values of black-body temperatures and power-law photon indices. If the spectral shape of the source does not depend on time (so $\alpha$ does not depend on time), the 6.4 keV intensity can be written as
\begin{equation}
\begin{split}
    F_{Fe}= 5.0 \times 10^{-30} \alpha \, \Omega\, N_{H_2} \, \left( \frac{\Delta t_{cloud}}{1.5 \text{yrs}}\right)^{-1}  
     \left(\frac{\delta_{Fe}}{2.95 \times 10^{-5}}\right) \\ \times \left(\frac{U_{2-10}\quad}{10^{42} \ \text{erg}} \right) \quad \text{[photons s}^{-1}\text{cm}^{-2}\text{]},
\end{split}
\end{equation}
where we assumed the cloud to be located at the GC ($D=8.3$ kpc), and Y=0.3.

\section{}

\begin{figure}[]
    \centering
    \includegraphics[width=0.9\linewidth]{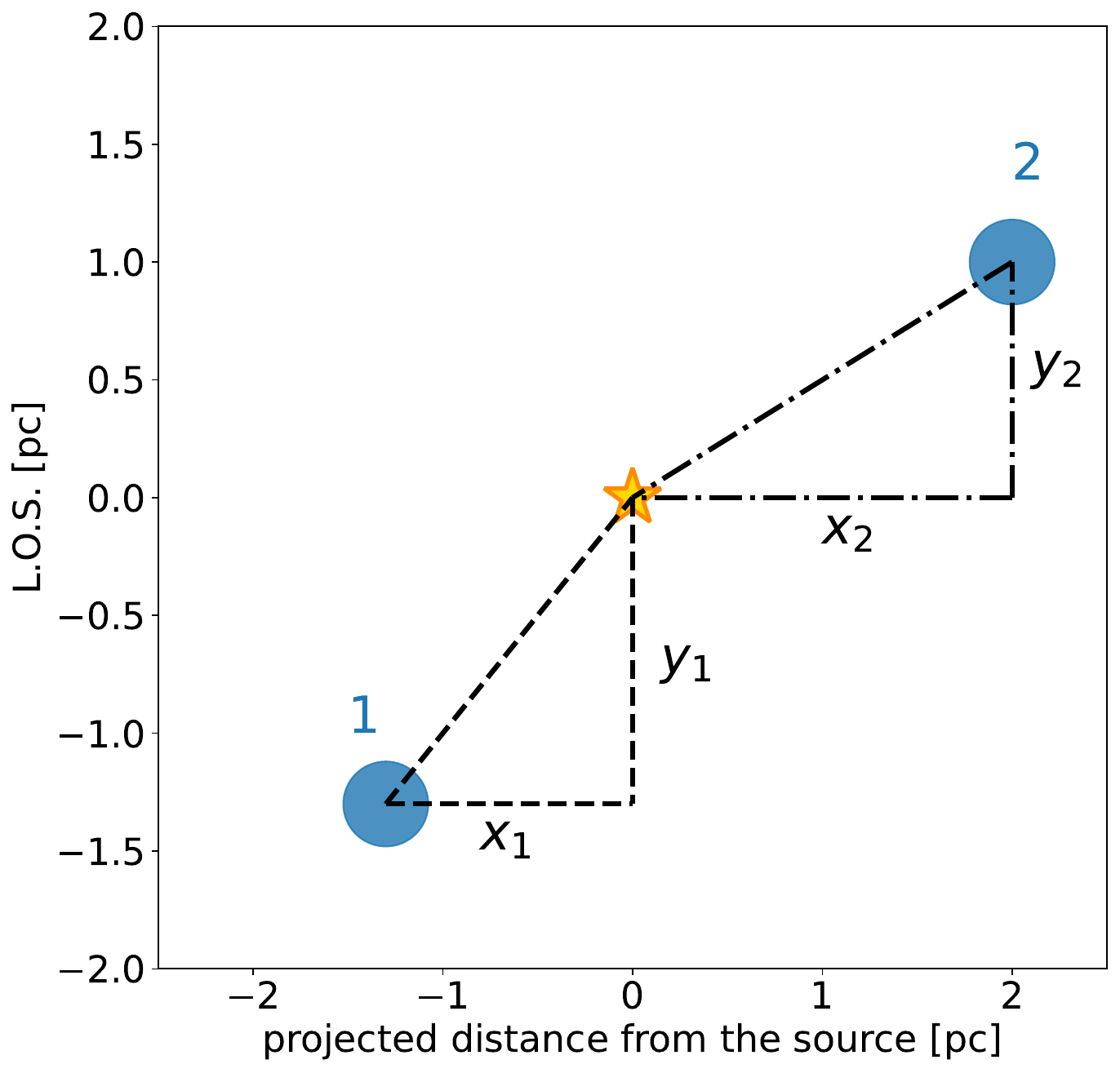}
    \caption{Light path difference for two molecular clouds. The illuminating source is located in (0.0, 0.0) pc. The observer is located at a (large) negative value along the LOS. The first cloud is in front of the source and the second behind. The x-axis is the projected distance in the plane of the sky.}
    \label{fig:distances}
\end{figure}

The time delay between the outburst of the illuminating source and the echo radiation can be used to derived their relative positions along the LOS. Fig.\ref{fig:distances} shows two different clouds, one in front and the other behind an illuminating source. The system is assumed to be far away from the observer. For the cloud in front of the source (cloud no. 1 in the figure), the time delay, $\Delta t_1$, is due to the longer path the light travels to reach the cloud and then the observer. The extra path the light travels, $\Delta l_1$, can be written as
\begin{equation}
\label{eq:time_delay1}
    \Delta l_1 = c \Delta t_1 = \sqrt{x_1^2+y_1^2}-y_1,
\end{equation}
where $x_1$ is the projected distance between the source and the cloud, and $y_1$ is the separation along the LOS. In all the considered cases, $x_1$ is derived from the measured angular separation ($\theta_1$) and the distance from the GC ($D=8.3$ kpc)\footnote{Formally, $x_1=\tan(\theta_1) \, (D-y_1)$, but in the GC case, $y_1$ can be neglected.}. $y_1$ can then be derived:
\begin{equation}
    y_1 =  \frac{x_1^2-(c \Delta t_1)^2}{2c \Delta t_1}
\end{equation}
Since $\Delta l_1 < x_1$, the time delay is less than the light-crossing time of the projected distance ($x_1$).

The projected wavefront that scans the cloud distribution has superluminal velocity, $v_1$. Indeed, from eq.\eqref{eq:time_delay1}:
\begin{equation}
    v_1=\frac{x_1}{\Delta t_1} = c \cot(\theta/2),
\end{equation}
where $\theta$ is the angle between the LOS and the line connecting the source with the cloud. Since the inclination angle of the CND is 70$^\circ$, $\theta=20^\circ$ in this specific case. The velocity is then $v_1=5.7 \, c$.

Analogously, if the cloud is behind the illuminating source (case no. 2  in the figure), the extra path the light travels is 
\begin{equation}
\label{eq:time_delay2}
    \Delta l_2 = c \Delta t_2 = \sqrt{x_2^2+y_2^2}+y_2,
\end{equation}
and the distance between the source and the cloud along the LOS becomes:
\begin{equation}
    y_2 =  -\frac{x_2^2-(c \Delta t_2)^2}{2c \Delta t_2}.
\end{equation}
In this case $\Delta l_2>x_2$, meaning that the time delay is greater than the light-crossing time of the projected distance.

The projected wavefront is subluminal. From eq.\eqref{eq:time_delay2}, the projected wavefront velocity is
\begin{equation}
    v_2=\frac{x_2}{\Delta t _2} = c \tan(\theta/2),
\end{equation}

where $\theta$ is the angle between the LOS and the line connecting the source with the cloud. We note that for the CND (inclination angle of 70$^\circ$) $\theta=20^\circ$ and $v_2= 0.18\, c$.

\end{appendix}
\end{document}